\documentclass[floatfix,amsmath,amssymb,aps,prb,groupedaddress,twocolumn,superscriptaddress]{revtex4-2}

\usepackage[margin=1.7cm]{geometry}

\usepackage{graphicx}
\usepackage{amsthm,amssymb,amsmath,braket,mathdots,mathtools}
\usepackage{bm}
\usepackage{CJK}

\usepackage{psfrag}
\usepackage{relsize,amsbsy}
\usepackage[export]{adjustbox}
\usepackage{dsfont}
\usepackage{xcolor,tikz}
\usepackage{lipsum}
\usepackage{ragged2e}

\usepackage{mwe}
\usepackage{tabularx}
\usepackage{multirow}
\usepackage[normalem]{ulem}
\usepackage[colorlinks,bookmarks=true,citecolor=blue,linkcolor=red,urlcolor=blue]{hyperref}

\DeclareMathOperator{\tr}{Tr}

\newcommand{\be}{\begin{align}}
\newcommand{\ee}{\end{align}}

\newcommand{\bit}{\begin{enumerate}}
\newcommand{\eit}{\end{enumerate}}

\begin{document}

\title{Real-time scattering in the lattice Schwinger model}

\author{Irene Papaefstathiou}
\affiliation{Max-Planck-Institut f{\"u}r Quantenoptik, Hans-Kopfermann-Str. 1, D-85748 Garching, Germany}
\affiliation{Munich Center for Quantum Science and Technology (MCQST), 80799 Munich, Germany}
\author{Johannes Knolle}
\affiliation{Department of Physics TQM, Technische Universität München, James-Franck-Straße 1, D-85748 Garching, Germany}
\affiliation{Munich Center for Quantum Science and Technology (MCQST), 80799 Munich, Germany}
\affiliation{Blackett Laboratory, Imperial College London, London SW7 2AZ, United Kingdom}
\author{Mari Carmen Ba\~nuls}
\affiliation{Max-Planck-Institut f{\"u}r Quantenoptik, Hans-Kopfermann-Str. 1, D-85748 Garching, Germany}
\affiliation{Munich Center for Quantum Science and Technology (MCQST), 80799 Munich, Germany}

\date{\today}

\begin{abstract}
   Tensor network methods have demonstrated their suitability for the study of equilibrium properties of lattice gauge theories, even close to the continuum limit. We use them in an out-of-equilibrium scenario, much less explored so far, by simulating the real-time collisions of composite mesons in the lattice Schwinger model. Constructing wave-packets of vector mesons at different incoming momenta, we observe the opening of the inelastic channel in which two heavier mesons are produced and identify the momentum threshold. To detect the products of the collision in the strong coupling regime we propose local quantitites that could be measured in current quantum simulation platforms.

\end{abstract}
\maketitle

\section{Introduction}

Scattering experiments are a well-established tool for probing fundamental physics. 
In particular, collision experiments allow the production of high energy and rare particles and thereby a study of their interactions.
Precise theoretical predictions of such processes, necessary for their interpretation, often involve contributions that can not be extracted from diagrammatic perturbation theory. This is the case, for instance, for hadron collisions, where non-perturbative effects of Quantum Chromodynamics (QCD) may play a significant role~\cite{Brambilla2014}.
The most powerful tool to address such non-perturbative regimes is lattice gauge theory (LGT), the discrete formulation of gauge field theories~\cite{Wilson1974}. Using advanced numerical methods, like Quantum Monte Carlo~\cite{foulkes2001,carlson2015}, LGT has allowed the successful exploration of strong coupling phenomena, such as the hadron spectra in QCD, but real time dynamics represents a challenge. Despite recent progress~\cite{Cichy2019}, a precise first-principles calculation of the scattering processes has not yet been possible, one of the reasons that motivates the search for alternative techniques~\cite{banuls2020ropp}.

In recent years quantum methods have unveiled potential alternative ways to explore fundamental physics (see~
\cite{banuls2020epj,davoudi2020,bass2021qtech,Funcke2023,Bauer2023, halimeh2023coldatom, dimeglio2023quantum} for reviews).
Their central focus are LGTs, 
which also appear as effective low-energy descriptions of strongly correlated condensed matter systems~\cite{kogut1979introduction}. The Hamiltonian version of LGTs then constitutes a 
natural object for quantum simulations. While a full simulation of QCD is still beyond reach, current efforts are focused on simplified and low-dimensional LGTs. An important development are the quantum-information based tensor network (TN) methods~\cite{Verstraete2008,SCHOLLWOCK201196,Orus2014annphys,Silvi2019tn,Okunishi2022,Banuls2023} which have been successfully applied to study spectral and thermal equilibrium properties of low dimensional LGTs~\cite{banuls2020ropp,felser2020u1,emonts2020z3,robaina2021z3,magnifico2021qed3d,emonts2022z2}, in 1+1D cases achieving some of the most precise existing extrapolations to the continuum.

Real time evolution phenomena, such as scattering dynamics, are among the most promising problems for a potential quantum advantage~\cite{tong2022provablyaccurate}, 
since Monte Carlo methods suffer in this case from a sign problem.
Tensor network algorithms, on the other hand, can be used to simulate LGT dynamics, but only  
for a limited time, before the entanglement in the system becomes too large for an efficient TN description. 
Yet, out-of-equilibrium simulations of LGT with TN have been used to explore the dynamics of pair production~\cite{buyens2014} and string breaking dynamics in several LGT models~\cite{kuehn2015string,buyens2016,pichler2016realt,chanda2020quenches,notarnicola2020ryd}.
Whereas far from the precision achieved by their equilibrium counterparts and not extrapolated to the continuum  limit, such simulations overcome the limits of classical Monte Carlo methods 
and constitute a valuable tool to prepare and analyze the potential quantum simulations.

More recently, scattering experiments in LGT have attracted the attention of TN calculations and quantum simulation proposals~\cite{pichler2016realt, rigobello2021entanglement, belyansky2023high}.
 The basic phenomenology of confinement~\cite{kormos2017real,vovrosh2022confinement} and meson scattering has also been explored recently in quantum spin models which share properties of LGT especially in 1+1D~\cite{vandamme2021scat,surace2021scattering,karpov2022spatiotemporal,milsted2022collisions, vovrosh2022hadronformation,chai2023thirring, farrell2024quantum}.

In this paper we use TN techniques to study elastic and inelastic scattering of composite particles in the lattice Schwinger model, which is the discrete lattice version of Quantum Electrodynamics in one spatial dimension.
Due to its similarities with more complex LGTs, including confinement and chiral symmetry breaking, 
the Schwinger model is a standard testbench for LGT methods, and has been focus of TN simulations, quantum simulation proposals and even experimental realizations (see~\cite{banuls2020epj,Shaw2020quantumalgorithms} for reviews). Studies of scattering processes in the Schwinger model  have been more scarce.

In absence of a background field, the spectrum of the model contains two stable particles, referred to as \emph{vector} and \emph{scalar}.
Elastic processes between two vector mesons, below the threshold of production of the scalar, have been simulated  in the strong coupling~\cite{pichler2016realt} and in the weak and intermediate coupling~\cite{rigobello2021entanglement} regimes. 
In this regime, the particles collide and can bounce back, without the creation of new particles post-collision, but generating entanglement between the mesons. A first approach to the inelastic regime was explored recently in~\cite{belyansky2023high}, with a focus on the thermodynamic limit and the non-perturbative regime near the confinement-deconfinement critical point, and using a bosonized formulation of the model, aimed at proposing a quantum simulation scheme
that could be implemented in circuit-QED.

In this work, we focus instead on the strong coupling regime at vanishing background field, and explore the energy threshold required to open the inelastic channel and obtain particle production after the collision.
Working in this 
regime allows us to propose simple initial state preparations and observables that could be amenable to realization on quantum simulators.

The paper is structured as follows: In Sec.~\ref{sec: Model} we briefly present the Schwinger model in the continuum limit and we show the lattice discretization we use in this work. In Sec.~\ref{sec: Methods} we discuss the preparation of the initial state of two meson wavepackets, indicate the momentum threshold for particle production and present the method used for the time evolution of the system. In Sec.~\ref{sec: Results} we show the results of meson-meson collisions. We specifically identify the energy threshold for particle production and we compare our results with the predicted threshold obtained with DMRG calculations.

\section{Model}\label{sec: Model}
 
The Schwinger model~\cite{schwinger1962gauge} is Quantum Electrodynamics (QED) in one spatial dimension. 
It is the simplest gauge theory including dynamical matter, and exactly solvable for massless fermions, yet it exhibits 
common features with those of Quantum Chromodynamics, like confinement and chiral symmetry breaking~\cite{coleman1976more}.
Its lattice discretization is commonly used as a benchmark for numerical LGT methods~\cite{hamer1982,hamer1997series}, and has in particular been 
widely adopted in studies of Tensor Network techniques for lattice gauge theories, 
which started~\cite{Byrnes:2002nv} with the application to the model of the Density Matrix Renormalization Group (DMRG) algorithm.

\subsection{The continuum limit} \label{sec: Model, continuum}

The continuum Hamiltonian of the Schwinger model in the temporal gauge can be written:
 \begin{equation}\label{eq: Schwingercontinuum}
    \begin{aligned}
    \mathcal{H}= &-\int dx \Big(i\bar{\Psi}(x)\gamma^{1}\big[\partial_{1}+ig A_{1}(x)\big]\Psi(x)\Big)\\
    &+\int dx \Big( m\bar{\Psi}(x)\Psi(x)+\frac{1}{2} E(x)^2\Big),
   \end{aligned}
\end{equation}
with  $\gamma^{0}$ and $\gamma^{1}$ the Dirac matrices in the temporal and spatial dimension, whereas $\partial_{1}$ is the partial derivative with respect to $x$.  The fermionic field $\Psi (x)$ has two components and its conjugate is $\bar{\Psi}=\Psi^{\dagger}\gamma^{0}$. $m$ is the fermionic mass and $g$ is the coupling constant. In an infinite volume, the only independent parameter of the model is the adimensional ratio $m/g$. Note that in the temporal gauge $A^{0}=0$. The electric field $E(x)$ is then related to the vector gauge field $A^{1}(x)$ as  $E(x)=-\partial_{0}A^{1}(x)$.

The continuum Schwinger model is exactly solvable for the case $m/g=0$ and $m/g\to\infty$~\cite{schwinger1, schwinger1962gauge}. For $m/g=0$, the Schwinger model can be solved using bosonization and corresponds to a free massive boson, the so-called Schwinger boson. For non-zero $m/g$, this boson becomes interacting.

We denote the ground state energy as $E_{\mathrm{GS}}$. The first stable particle of the theory is the vector meson, a bound state (Schwinger boson) with mass $M_{\mathrm{V}}$. The second stable particle is the scalar meson, with mass $M_{\mathrm{S}}$, which can be considered a bound state of two Schwinger bosons~\cite{coleman1976more,adam1997}. 

These two bound states are stable particles with parity and charge conjugation being well-defined quantum numbers (CP symmetry)~\cite{coleman1976more}.
The bound state of two Schwinger bosons belongs in the same sector as the ground state (scalar sector), whereas the Schwinger boson belongs to the vector sector.

In open boundary conditions, integrating Gauss law fixes the electric field up to a constant, which can be interpreted as an external background field $g\theta/2\pi$, with $\theta$ the so-called vacuum angle~\cite{coleman75,coleman1976more}. In the strong coupling (small mass) regime we target in this study, it is possible to estimate the particle masses perturbatively. Specifically, the mass of the mesons $M_{\mathrm{V}}$, in the strong coupling regime can be obtained from second-order perturbation theory as~\cite{ADAM1996383, ADAM2003}:
\begin{align}
\label{eq: meson mass}
    M_{\mathrm{V}}^{2}&\equiv \mu_{2}^{2} 
    \\&=\mu_{0}^{2} \bigg(1+3.5621 \frac{m}{\mu_{0}} \cos(\theta) \notag
    \\&+5.4807\frac{m^{2}}{\mu_{0}^{2}}-2.0933 \frac{m^{2}}{\mu_{0}^{2}}\cos(2\theta)\bigg)
    \notag
\end{align}
with $\mu_{0}^{2}=\frac{g^{2}}{\pi}$ the result from zero-order perturbation theory. 
The mass of the second excitation of the theory, the bound state $M_{\mathrm{S}}$, is given in second-order perturbation theory by~\cite{adam1997}:
\begin{align}
\label{eq: BS mass}
    M_{\mathrm{S}}^{2}&=4M_{\mathrm{V}}^{2}-\Delta
\end{align}
with $\Delta=\frac{4\pi^{4}m^{2}\Sigma^{2}\cos^{2}(\theta)}{M_{\mathrm{V}}^{2}}$. $\Sigma=\frac{g^{\gamma}}{2\pi}\mu_{0}$ is the fermion condensate with $\gamma$ the Euler constant~\cite{ADAM1996383}.

The form of the wavefuction of the bound states in the strong coupling regime of the continuum limit was previously studied first by Ref.~\cite{mo1993basis} up to and including four-body states and then by Ref.~\cite{harada1995} up to and including six-body states with the use of the six-body light-front Tamm-Dancoff approximation. In these works it was found that the composition of the bound states depends on the fermionic mass $m/g$. They specifically consist of a two-body component, a four-body component and a negligible six-body component. The contribution of the four-body component increases as the fermionic mass $m/g$ is getting smaller.

\subsection{The lattice}
This section presents the lattice Schwinger model, the scattering processes of which will be studied in this paper. Here we are using the Kogut-Susskind staggered fermion formulation for the model \cite{kogut1975j}:
\begin{align}
    W &=\frac{g^{2}a}{2}\sum_{n}L_{n}^{2}+ m\sum_{n}(-1)^{n}\Phi^{\dagger}_{n}\Phi_{n},\nonumber\\
    &-\frac{i}{2a} \sum_{n}\left(\Phi_{n}^{\dagger}e^{i\theta_{n}}\Phi_{n+1}-h.c.\right)\,,
\label{eq: Schwinger lattice}
\end{align}
with the lattice spacing being $ga$ in coupling units. Specifically, we focus on the strong coupling regime, which means $m/g \ll 1 $.

The fermionic annihilation(creation) operators ${\Phi}_{n}({\Phi}_{n}^{\dagger})$ reside on the lattice sites $n$. As single-component fermionic operators, they satisfy canonical anticommutation relations $\{\Phi^{\dagger}_{n}, \Phi_{m} \}=\delta_{nm}$ and
$\{\Phi_{n}, \Phi_{m} \}=0$.
The gauge fields $\theta_n$ and $L_{n}$ are the lattice equivalent to the vector potential and the electric field in the continuum. They occupy the links between two nearest neighbouring sites ($n$ and $n+1$). They are canonically conjugate operators, satisfying the commutation relation $[\theta_{n},L_{m}]=i\delta_{nm}$.
In addition, the lattice version of the Gauss law reads~\cite{hamer1997series}:
\begin{align}
\label{gauss law}
 L_{n}-L_{n-1}=\Phi^{\dagger}_{n}\Phi_{n}-\frac{1}{2}\left[1-(-1)^{n}  \right]   
\end{align}
and needs to be satisfied by the physical states of the system.

By a Jordan-Wigner transformation, the Hamiltonian~\eqref{eq: Schwinger lattice} can be mapped onto a spin model. The Gauss law, which now reads:
\begin{align}
\label{eq: gauss law spin}
L_{n}-L_{n-1}&=\frac{1}{2} \big( (-1)^{n}+(\sigma_{n}^{z}) \big),
\end{align}
can be integrated out if we consider open boundary conditions, resulting in a spin Hamiltonian with long range interactions~\cite{banks1976strong,hamer1997series}.
After rescaling it with a global $2/(ag^2)$ factor, one is left with the adimensional Hamiltonian:
 \begin{align}
    H &=\frac{2}{g^2 a}W= H_{x}+H_{\mu}+H_{l} \nonumber \\
    &=x\sum_{n=0}^{N-2}\left[\sigma_{n}^{+}\sigma_{n+1}^{-}+\sigma_{n}^{-}\sigma_{n+1}^{+}\right]\nonumber\\
    &+\frac{\mu}{2}\sum_{n=0}^{N-1}\left[1+(-1)^{n}\sigma^{z}_{n}\right]\nonumber\\ 
    &+\sum_{n=0}^{N-2}\left[ l+\frac{1}{2}\sum_{k=0}^{n}((-1)^{k}+\sigma_{k}^{z}) \right]^{2},
    \label{eq: Schwinger spin}
\end{align}
with $x=\frac{1}{g^{2}a^{2}}$ 
and $\mu=\frac{2m}{g^{2}a}$ being adimensional parameters and $\sigma^{\pm}=\frac{1}{2}(\sigma^{x}\pm i\sigma^{y})$. $l$ is the background electric field and is related to the vacuum angle as following: $l=\theta/2\pi.$ It is set to zero in this work, $l=0$.

 The lattice model with open boundary conditions breaks most of the continuum symmetries. In particular, charge conjugation is no longer a good quantum number. However, the symmetry is restored in the thermodynamic limit and, if the system is large enough, it is possible to define an approximate charge conjugation operator whose expectation value can be used to distinguish the scalar and vector sectors~\cite{banuls2013mass}.
 
\subsubsection{The strong coupling regime}

In this paper we study the spin formulation of the lattice Schwinger model solely focusing on the strong coupling regime. Since here we use the rescaled Hamiltonian of Eq.~\eqref{eq: Schwinger spin} with the adimensional parameters $\mu$ and $x$, being in the strong coupling regime for a finite lattice means $x \ll 1$. In this regime $H_{x}$ can be considered the perturbation, with $H_{\mu}+H_{l}$ being the unperturbed part of the Hamiltonian. 
The strong coupling regime offers the advantage of having a simple Ansatz for the initial mesons of the collision, as well as a clear energy threshold for producing the scalar mesons of interest. We note that, a truncated Hilbert space approach, which has been employed for spin chains~\cite{vovrosh2022hadronformation}, would not work here. 

In the infinite coupling limit, $x=0$, the ground state of the unperturbed Hamiltonian $H_{\mu}+H_{l}$ is the state with all the sites empty:
\begin{align}
    |0\rangle &= |\downarrow \uparrow \downarrow \uparrow \cdots \downarrow \uparrow \rangle
    \label{eq: ground state, infinite coupling}
\end{align}
The sites are numbered from left to right $n=0, 1, \dots, N-1$. Due to the fact that we are using the staggered fermions formulation, the odd sites correspond to the fermions and the even sites to the antifermions. Specifically, the occupied sites are $ |\downarrow \rangle_{O}$ for the odd sites and $|\uparrow \rangle_{E}$ for the even sites ($|\uparrow \rangle_{O}$ and $|\downarrow \rangle_{E}$ are the empty odd and even sites correspondingly). Therefore, $ |\downarrow \rangle_{O}$ are the particles (electrons) and $|\uparrow \rangle_{E}$ are the anti-particles (positrons).

In the limit of infinite coupling, $x=0$, the first excitation of the theory $|1_{\mathrm{V}} \rangle$ (the vector meson), with energy $E_{\mathrm{V}}$, can be obtained simply as:
\begin{align}
    |1_{V} \rangle &= \frac{1}{\sqrt{N}} \sum_{n} \big( \sigma^{+}_{n}\sigma^{-}_{n+1}- \sigma^{+}_{n+1}\sigma^{-}_{n} \big) |0 \rangle.
    \label{eq: meson}
\end{align}
For small but non-vanishing values of $x$, the strong coupling expansion (SCE)~\cite{hamer1997series}, a perturbative expansion in $x$,  can be used to obtain very precise estimates of the 
ground state energy and the mass spectrum, for small bare fermion mass~\cite{CICHY20131666}.

Our goal is to create states of the mesons with well defined momenta. Although we do not have the explicit solution for the operator that creates these particles, in the strong coupling regime we can use an Ansatz that resembles the first order  of SCE~\eqref{eq: meson}, but where the strong coupling vacuum has been substituted by the true ground state at finite $x$,  which we denote $|\Omega \rangle$ and compute numerically following the method in~\cite{banuls2013mass}. Therefore, our Ansatz for the state of the mesons can be written as:
\begin{align}
    |1_{V} \rangle &= \frac{1}{\sqrt{N}} \sum_{n} \big( \sigma^{+}_{n}\sigma^{-}_{n+1}- \sigma^{+}_{n+1}\sigma^{-}_{n} \big) |\Omega  \rangle.
    \label{eq: meson}
\end{align}

\section{Methods}\label{sec: Methods}
\subsection{Preparing the initial state}
\label{sec: Methods--initial state}
The first step for studying the elastic and inelastic meson-meson collisions is creating two meson wavepackets in the initial state and giving them opposite momentum so that they eventually collide.
We prepare two wavepackets for the mesons as Gaussian superpositions of the operator in Eq.~\eqref{eq: meson} acting on different positions, 
namely:
\begin{align}
 \label{eq: wavepackets}
    |\phi_{i}\rangle &= \mathcal{N}_{i} \sum_{n=\alpha^{\mathrm{beg}}_{i}}^{\alpha^{\mathrm{end}}_{i}} e^{-(n-c_{i})^{2}/(2\sigma^{2})}e^{-ink_{i}} \big( \sigma^{+}_{n}\sigma^{-}_{n+1}-\sigma^{+}_{n+1}\sigma^{-}_{n} \big) |\Omega \rangle  \notag\\
   &= \mathcal{O}_{i} |\Omega \rangle \notag \\
\end{align}
with indices $i=1,\ 2$ corresponding to the two mesons. We choose the wavepackets to have centers $c_i$, and common width $\sigma$. Initially, the wavepacket of the first (second) meson is spread in the left (right) half of the system. We therefore have $\alpha_{1}^{\mathrm{beg}}=0$, $\alpha_{1}^{\mathrm{end}}=N/2-1$ for the first meson and $\alpha_{2}^{\mathrm{beg}}=N/2$ and $\alpha_{2}^{\mathrm{end}}=N-1$ for the second one.
We consider the mesons to have opposite momenta with the same magnitude, $k_{1}=-k_{2}=k$.
$\mathcal{N}_{i}$ are the necessary normalization factors.  
The total initial state is constructed as the product:
\begin{align}
    |\phi \rangle &= ( \mathcal{O}_{1} \mathcal{O}_{2} ) |\Omega \rangle.
    \label{eq: initial state with O operators}
\end{align}

Notice that this approach is made possible by working in the strong coupling regime, as this provides us with the explicit approximate description of the vector meson state shown in Eq.~\eqref{eq: meson}, which can be directly used for the creation of the wavepackets.

\subsection{Momentum threshold $k_{\mathrm{thr}}$}
\label{sec: Methods--momentum threshold}

In order to open an inelastic channel (i.e. to produce a final state allowed by the symmetries of the problem), the momentum of the incoming wavepackets needs to be high enough  to produce the new particles satisfying the kinematic constraints imposed by energy and momentum conservation. 

We consider collisions in the center of mass of the incoming mesons, where the initial wavepackets have opposite spatial momenta $p_{1}=-p_{2}\equiv {p}$. Momentum conservation imposes that the outgoing particles, two identical scalar mesons, have the same momenta with opposite directions, $q$ and $-q$.
Conservation of energy then requires:
\begin{equation}
\label{eq: energy conservation}
2\sqrt{M_\mathrm{V}^{2}+p^2}=2\sqrt{M_\mathrm{S}^2+q^2}
\end{equation}
The final state with the smallest possible energy would correspond to creating the pair of scalars at rest, $q=0$. This sets the momentum threshold for the incoming vector mesons at:
\begin{align}
\label{eq: threshold}
 p^{2}&=p_{\mathrm{thr}}^{2}=M_{\mathrm{S}}^{2}-M_{\mathrm{V}}^{2}.   
\end{align}

Notice that the relations above are written in terms of physical (continuum) momenta. Our initial state~\eqref{eq: wavepackets} is however written in terms of adimensional lattice quantities. In particular, the lattice momentum $k$ is related to the physical one as $k=p/(g\sqrt{x})$. 

Because we work on a finite lattice with open boundary conditions, the scalar mesons cannot be produced at rest, but will carry at least the smallest lattice momentum $k_{\min}=O(1/N)$, which will correct the threshold.

In the continuum limit, one could calculate the momentum threshold with the use of perturbation theory and Eq.~\eqref{eq: meson mass} and Eq.~\eqref{eq: BS mass}. Since in our parameter regime, $N=100$, $x=1$ and $\mu=2\times 10^{-5}$, we are far away from that limit, we instead calculate the masses using a variational optimization of MPS, with the algorithm in ref.~\cite{banuls2013mass} for which we obtain the masses $M_V/g=0.785$, $M_S/g=1.370$. With these values we estimate the lattice momentum threshold at:
\begin{align} \label{eq: threshold DMRG}
 k_{\mathrm{thr}}^{\mathrm{MPS}}&\approx 1.123.   
\end{align}

\subsection{Time Evolution}

In our work, we use TN methods to prepare the initial state and simulate the time evolution. Firstly, we find a matrix product state~\cite{SCHOLLWOCK201196} approximation to the interacting vacuum $|\Omega \rangle$ by optimizing variationally the MPS ansatz  that minimizes the energy, using the method in~\cite{banuls2013mass}. We truncate the bond dimension of the ground state to $D_{\mathrm{GS}}=40$. The operators $\mathcal{O}_{1}$ and $\mathcal{O}_{2}$ of Eq.\eqref{eq: wavepackets} can be written as exact matrix product operators (MPOs)~\cite{Pirvu_2010} with bond dimension $D=4$. They can be then applied to the ground state, $|\Omega \rangle$ to create the meson wavepackets of Eq.~\eqref{eq: wavepackets}. We truncate the bond dimension of the wavepackets to $D_{\mathrm{V}}=50$, which, when compared to  the exact initial state of meson wavepackets (with bond dimension $4D_{\mathrm{GS}}$), has fidelity very close to 1 (with negligible corrections).

After creating the initial state with the two meson wavepackets, we perform time evolution with standard tensor network techniques~\cite{Verstraete2008,SCHOLLWOCK201196}. Specifically, we use a second order Suzuki-Trotter expansion of the evolution operator~\cite{Trotter1959OnTP, Suzuki1990FractalDO} with time step $\delta$, using the splitting of the Hamiltonian described in~\cite{thermal2015}, namely:
\begin{align}
    e^{-i\tau H} &\approx \bigg( e^{-i\frac{\delta}{2}H_{x}^{e}}e^{-i\frac{\delta}{2}H_{z}} e^{-i\delta H_{x}^{o}} e^{-i\frac{\delta}{2}H_{z}} e^{-i\frac{\delta}{2}H_{x}^{e}} \bigg)^{\tau/\delta}.
    \label{eq:trotter}
\end{align}
 $H_{x}^{e}$ and $H_{x}^{o}$ are the hopping terms (part of $H_{x}$) acting on the two-body terms of even-odd and odd-even pairs of sites correspondingly. $H_{z}$ is the mass term of the Hamiltonian together with the electric field term, $H_{z}=H_{\mu}+H_{l}$.

The exponentials $e^{-i\frac{\delta H_{x}^{e}}{2}}$ and $e^{i\frac{-\delta H_{x}^{o}}{2}}$ can be written as exact MPOs with bond dimension 4. The exponential of the term $H_{z}$ can also be written as an exact MPO, diagonal in the $z$ basis, with bond dimension that scales linearly with the system size as $\chi=N+1$~\cite{thermal2015}.

Specifically, the elements of the MPO are:
\begin{align}
(M_{n}^{ij})_{L_{n-1}L_{n}}=e^{-i \delta h_{n}(i)}\delta_{ij} \delta_{L_n-L_{n-1}, G_n(i)},
\end{align}
where the virtual indices label the electric flux on the links, $L_n\in[-N/2,N/2]$ and the second delta ensures the fulfilment of Gauss law~\eqref{eq: gauss law spin},
\begin{align}
G_n(i)=\frac{1}{2}[(-1)^n+(-1)^i],\quad i=0,1,
\end{align}
and $h_{n}(i)=\mu (-1)^n G_n(i)+L_{n}^{2}$ for $n<N-1$, with $h_{N-1}(i)=\mu (-1)^{N-1} G_{N-1}(i)$.

To be able to deal with large system sizes (of the order of 100 or more sites) it is convenient to truncate the bond dimension of this MPO to a constant value $\chi_{\mathrm{tr}}$, which can be related to a cutoff in the absolute value of the electric flux $L_{\max}$, as $\chi_{\mathrm{tr}}=2 L_{\max}+1$. This truncation, which was successfully used for thermal equilibrium states in~\cite{thermal2015,chiral2016}, will be also valid in our case as far as the dynamics does not generate states with large occupations of the electric flux. The cutoff in the electric flux is restricted to $L_{\max}=8$ in our study. Applying the evolution operator increases the bond dimension of the time-evolved state $|\Psi (t) \rangle$ at each time step. In order to control the computational cost, we truncate the bond dimension of the evolved state to a maximum value of $D=50$. Even though  we use relatively small bond dimensions for the ground state, the initial mesons and the evolved state, as well as a relatively small cutoff of the electric flux, we check that our qualitative observations are stable under changes of the truncation parameters, and this level of precision suffices to observe the effects of inelastic scattering in the cases under study. Specifically, knowing that the total energy has to be conserved in our system, we control the errors by measuring the (relative) change of the total energy in comparison to the initial total energy of the system. We find that for the times of interest (up to $gt_{\mathrm{phys}}=140$) the relative difference is less than $6\%$.
 
\begin{figure}[t!]
\centering
\includegraphics[width=0.9\columnwidth]{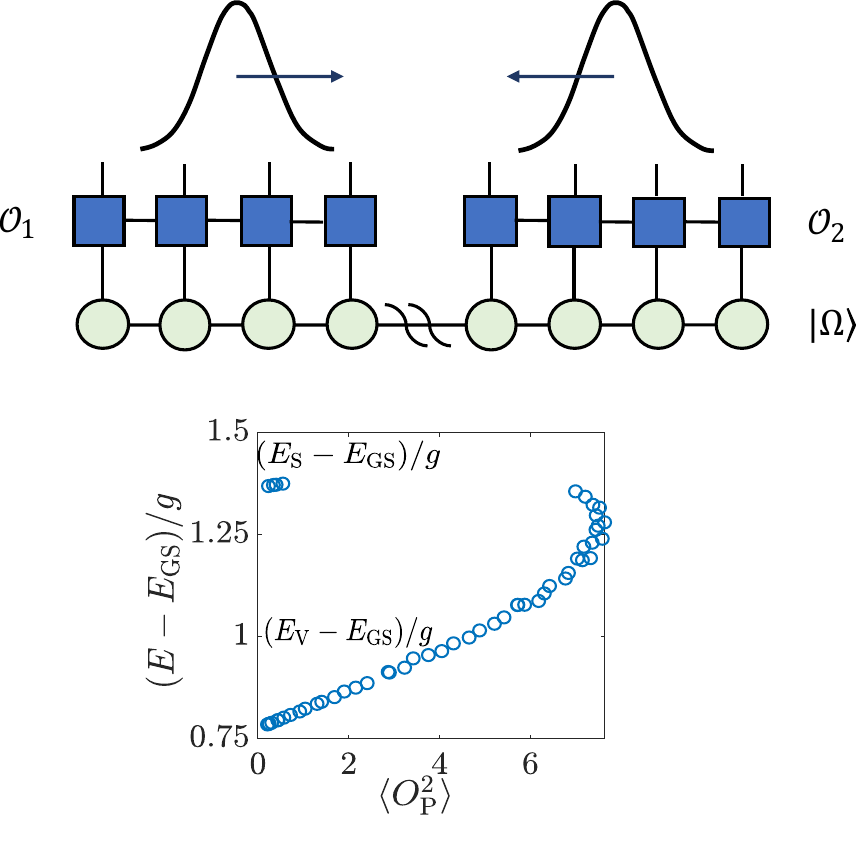}
\caption{The upper panel shows the initial state preparation following Eq.~\eqref{eq: initial state with O operators}. The lower panel shows the dispersion relation of the adimensional Hamiltonian of Eq.~\eqref{eq: Schwinger spin} as
$E=f(\langle O_{P}^{2} \rangle)$, with $O_{P}=-ix\sum_{n}(\sigma_{n}^{-}\sigma_{n+1}^{z} \sigma_{n+2}^{+}-h.c.)$ the dimensionless momentum operator for the fermion field (see~\cite{banuls2013mass}).
The lower branch corresponds to the vector meson and the upper one to the scalar, for which only the lowest momentum excitations are shown. The ground state energy, $E_{\mathrm{GS}}/g$, is subtracted from all the energies plotted. We consider a finite system with lattice size $N=100$ and physical volume $N /\sqrt{x} =100$. The mass is $m/g = 10^{-5}$. The energies were calculated with variational MPS as in~\cite{banuls2013mass} with small bond dimension, $D=50$.}
\label{fig: mesons initial state preparation}
\end{figure}
 
\section{Results}\label{sec: Results}

In this section we present the results of elastic and inelastic meson-meson scattering processes. Specifically, in subsection~\ref{sec: entropy and electric field energy} we observe the first signs of particle production, when having energies above the momentum threshold of Eq.~\eqref{eq: threshold DMRG} (inelastic channel). In subsection~\ref{sec: energy correlators} we calculate quantities that can be directly measured in experiments, indicating the signal of particle production for the inelastic channel. In subsection~\ref{sec: bound states structure} we study the four-body projector. 

The results shown in this section correspond to a fixed fermion mass $m/g=10^{-5}$ and physical volume $N/\sqrt{x}=100$ (notice that we express all magnitudes in units of the coupling $g$).

These magnitudes are related to their corresponding lattice analogues through the  lattice spacing $ga=1/\sqrt{x}$.
In particular, the time parameter $\tau$ appearing in the evolution operator $e^{-iH\tau}$ \eqref{eq:trotter} is the adimensional lattice time. It is related to the physical time $g t_{\mathrm{phys}}$ as:
\begin{align}
    \tau &= \frac{gt_{\mathrm{phys}}}{2\sqrt{x}}
    \label{eq: time scaling}
\end{align}

The lattice size was fixed to $N=100$ sites, and the inverse lattice spacing to the value $x=1$ in all the cases below.

\begin{figure*} [t!]
\centering
\includegraphics[width=2\columnwidth]{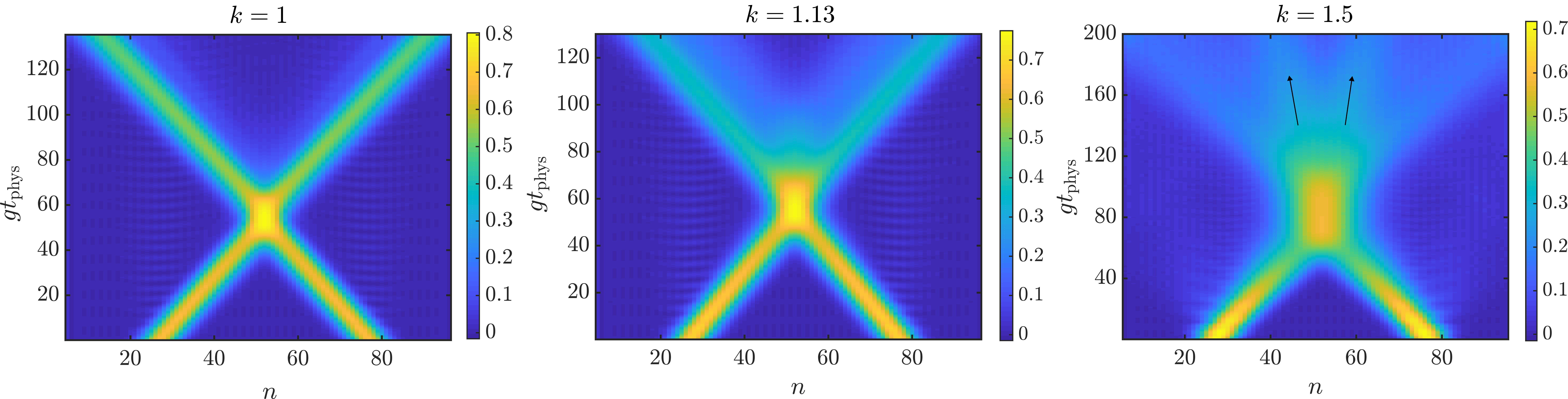}
\caption{ The plots show the two-site entanglement entropy $S(n,t)$, defined in the  main text~\eqref{eq:ent_entropy}, with the entanglement entropy of the vacuum subtracted from all the plots. The fermion mass is fixed to $m/g=10^{-5}$.  We consider a finite system with lattice size $N=100$ and physical volume $N/\sqrt{x} =100$. 
The standard deviation of the wavepackets is $\sigma=4$. In the left plot, the momentum magnitude of the incoming mesons is $k=1$ and only the elastic channel is present. The elastic channel corresponds to two mesons exiting the collision (yellow colour) with the same momentum as the momentum of the incoming mesons. In the middle plot, the momentum of the incoming mesons is $k=1.13$, slightly above the threshold predicted by the variational MPS calculation $k_{\mathrm{thr}}^{\mathrm{MPS}}$ ~\eqref{eq: threshold DMRG}. At this point, the inelastic channel opens, with two bound states produced with velocities close to 0. The signal of the two bound states that were created after the collision is visible a bit below $gt_{\mathrm{phys}} \sim 80$, with the two outgoing mesons coming from the elastic channel also being visible. The momentum of the incoming mesons for the plot on the right is $k=1.5$, also above the momentum threshold. Both the elastic and inelastic channels are open. The two bound states created from the collision (inelastic channel) have non-zero velocities with opposite signs and less magnitude than the momenta of the outgoing mesons of the elastic channel. The elastic channel is represented by the light blue cone-like shape, after the collision, whereas the velocities of the bound states are shown with the two arrows. 
The parameters used in the numerical simulation were bond dimension $D=50$, Trotter step $\delta=0.1$ and maximum electric flux $L_{trunc}=8$.}
\label{entanglement}
\end{figure*}

\subsection{Entropy}
\label{sec: entropy and electric field energy}

We first calculate the entanglement entropy $S(n,t)$ 
corresponding to a bipartition where one subsystem includes two neighboring sites $n$ and $n+1$, with $n=0,\dots N-2$, and  the other subsystem consists of the remaining  sites:
\begin{align}
    S(n, t)&= - \tr [\rho_{n,n+1}(t)\log_{2}\rho_{n,n+1} (t)],
    \label{eq:ent_entropy}
\end{align}
with $\rho_{n,n+1}$ being the reduced density matrix of one of the subsystems. $S(n,t)$ is interpreted as the entanglement between the sites $\{n,n+1\}$ and the rest of the system. 
Since the two-site reduced density matrix can in principle be recovered from local measurements, as recently realized in optical lattices in~\cite{impertro2023local},
this quantity could be more easily accessible in an experiment than the half-chain entanglement that has been analyzed in other scattering simulations~\cite{pichler2016realt,rigobello2021entanglement,belyansky2023high}. 

The results of the calculation of the entanglement entropy can be seen in Fig.~\ref{entanglement} and in Fig.~\ref{entanglement entropy, after collision}. We study three different cases depending on the magnitude of the momenta of the incoming mesons. The three cases consider momenta below, slightly above and above the momentum threshold $k_{\mathrm{thr}}^{\mathrm{MPS}}$. The first case is the case where the momenta are below the momentum threshold, as seen in the left plot of Fig.~\ref{entanglement}. We specifically have lattice momenta $k_{1}=-k_{2}=k=1$. Here, the incoming mesons do not have enough energy to create new particles. Therefore, only the elastic channel is present and the outgoing particles are two vector mesons with the same momenta as the incoming ones. Fig.~\ref{entanglement entropy, after collision} shows the excess entanglement entropy with respect to the vacuum after the collision, for physical time $gt_{\mathrm{phys}}=90$ (blue line). 
The two peaks represent the wavepackets of the two outgoing mesons. 
Notice the small increase in the entanglement between the two wavepackets in comparison to the entanglement entropy of the vacuum. Because the local entropy we compute is not directly measuring entanglement between the outgoing mesons, we attribute this increment to the fluctuations imprinted onto the vacuum by the collision.

\begin{figure}[h!]
\centering
\includegraphics[width=0.8\columnwidth]{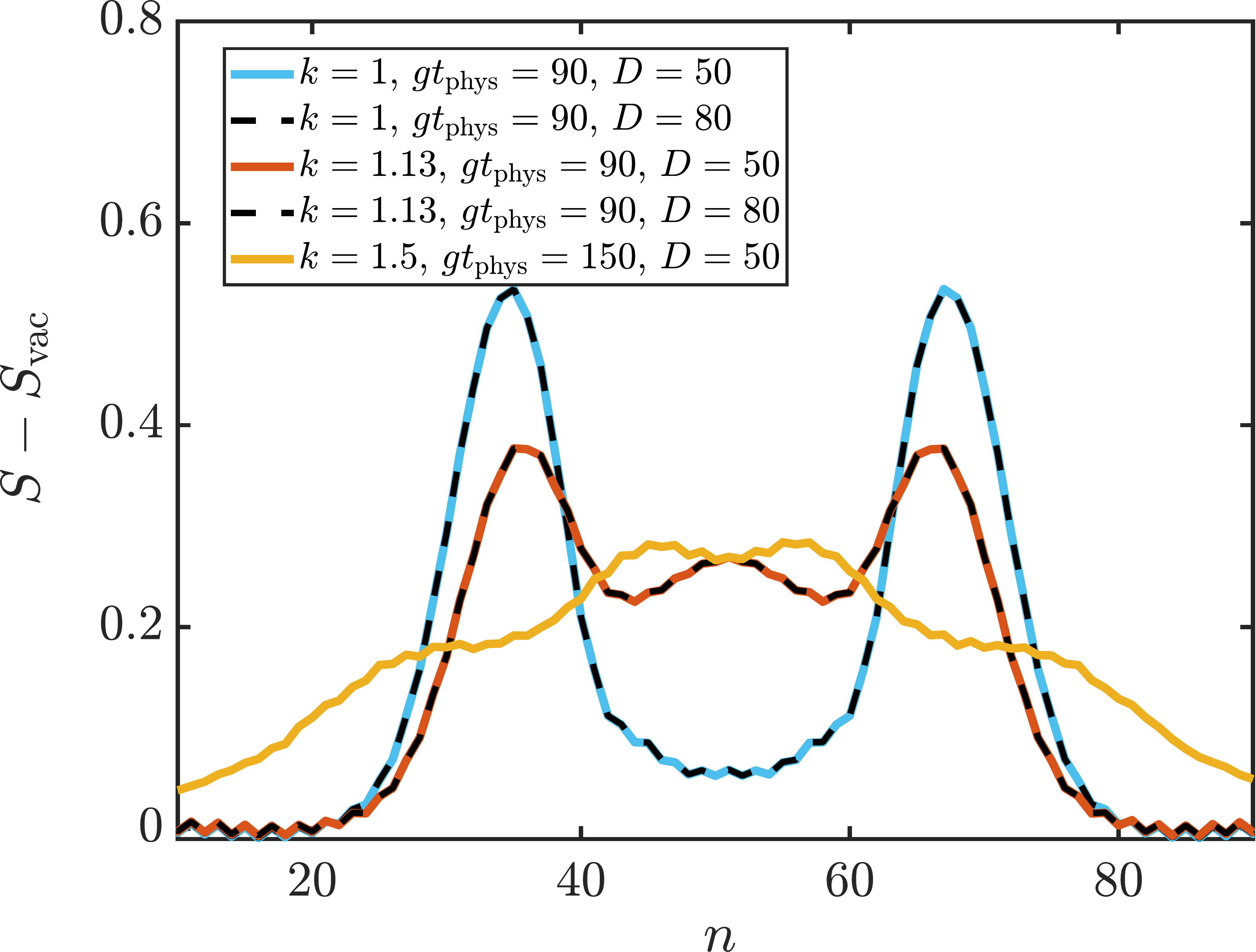}
\caption{The plot shows the two-site entanglement entropy $S(n,t)$ ~\eqref{eq:ent_entropy} minus the entanglement entropy of the vacuum, $S_{vac}$,  on a lattice with size $N=100$ and physical volume $N /\sqrt{x}=100$, for the cases where the initial particles have momenta below, slightly above and above the momentum threshold, at $k=1$, $k= 1.13$ and $k=1.5$.  The mass is $m/g=10^{-5}$, and the standard deviation of the wavepackets $\sigma=4$. 
As in Fig.~\ref{entanglement}, the parameters used in the numerical simulation were bond dimension $D=50$, Trotter step $\delta=0.1$ and maximum electric flux $L_{trunc}=8$. To demonstrate the numerical convergence, for the cases $k=1$ and $k=1.13$ we show also the results for $D=80$ (black dash lines) while keeping all the other parameters the same.}
\label{entanglement entropy, after collision}
\end{figure}

The second case is the middle plot of Fig.~\ref{entanglement}, where the incoming particles have enough energy to produce two scalar mesons of mass $M_{S}$ with zero velocity. 
Specifically, as mentioned in Sec.~\ref{sec: Methods--initial state} in Eq.~\ref{eq: threshold DMRG}, the threshold for particle production is at $k_{\mathrm{thr}}^{\mathrm{MPS}}\approx 1.12$. As the initial momenta of the vector mesons is $k=1.13$, we expect producing scalar mesons that are almost at rest after the collision. Notice here that it is not possible to create particles with momentum exactly 0, because as mentioned in Sec.~\ref{sec: Methods--momentum threshold}, the scalar mesons have to have at least momentum $k_{\min}=O(1/N)$.
One could observe that after the collision ($ t_{\mathrm{phys}} \sim 80$) there is a sign of particle production in the middle of the system, where the entanglement entropy is greater than the entanglement entropy between the outgoing particles of the elastic channel. This signal can be more clearly observed in Fig.~\ref{entanglement entropy, after collision} (orange line). The peak in the middle of the system corresponds to the two bound states produced with almost zero velocities. 
The higher peaks on the left and right of this signal correspond to the outgoing mesons of the elastic channel.  

Finally, the right plot of Fig.~\ref{entanglement} represents the case where the momentum of the incoming particles is considerably larger than the threshold, such that they have enough energy to produce two scalar particles of mass $M_{S}$, and the excess is transformed in kinetic energy of the outgoing mesons. In this case, we can notice both the elastic and the inelastic channels.
Since the scalar mesons are produced with small velocities, their signal is found near the middle of the system, as also seen by Fig~\ref{entanglement entropy, after collision}. One could also observe the signal of the elastic channel, near the edges of the same plot.

We notice here that the momentum threshold for particle production calculated from the variational MPS spectrum calculations was $k_{\mathrm{thr}}^{\mathrm{MPS}}\approx 1.12$. The simulated collisions behave as expected, with the first particle production observed at $k=1.13$, and no signs of particle production at $k=1$. We could also compare with the momentum threshold predicted by perturbation theory in the continuum limit, $p_{\mathrm{thr}}^{\mathrm{continuum}}/g\approx 0.98$, which was done with the use of Eq.~\eqref{eq: threshold}, Eq.~\eqref{eq: meson mass} and Eq.~\eqref{eq: BS mass}.

\subsection{Correlators}
\label{sec: energy correlators}
\begin{figure}[h!]
\centering
\includegraphics[width=0.85\columnwidth]{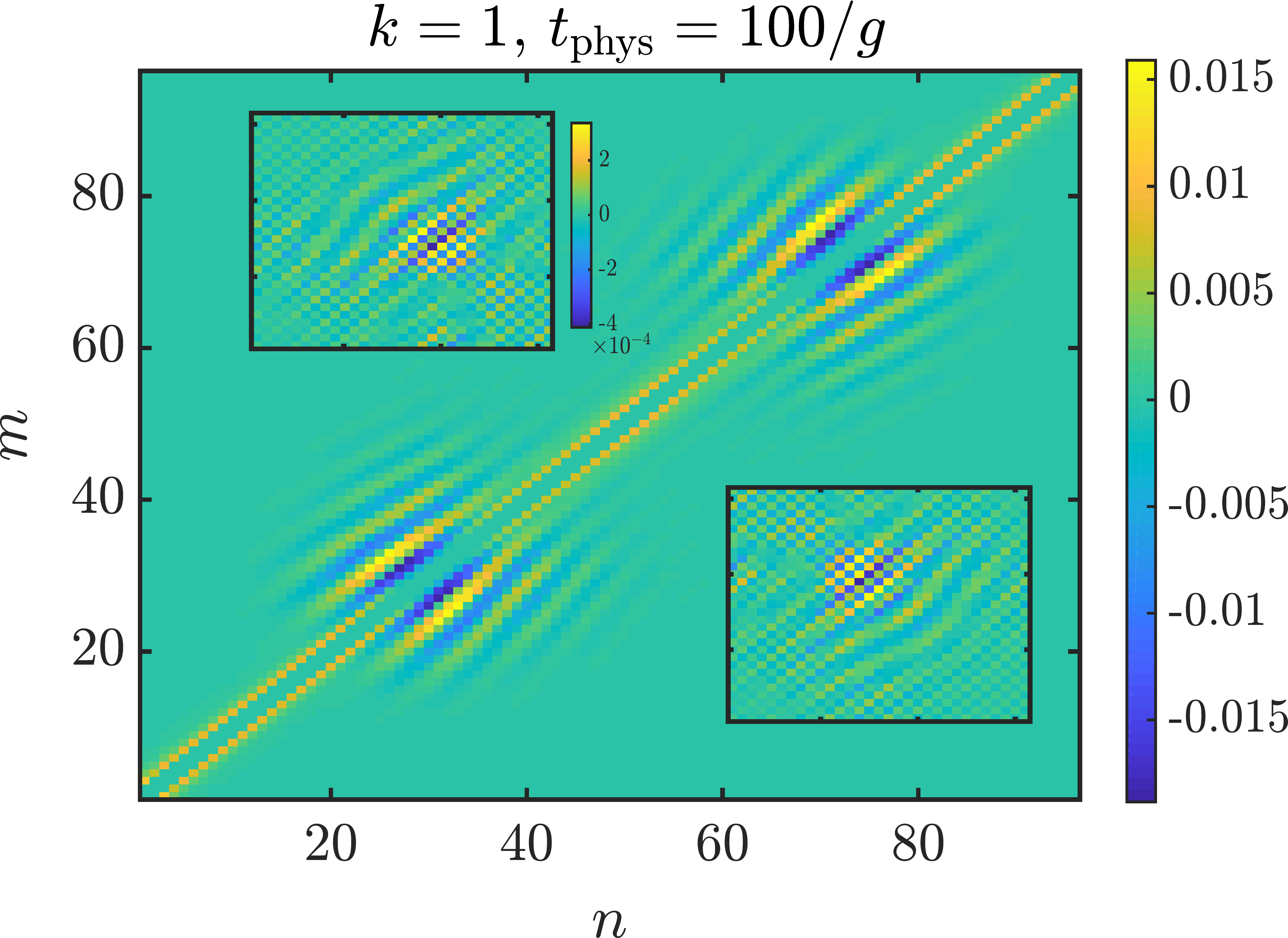}
\caption{This plot shows the electric flux correlator, $C^{mn}_{l}$, of a system size $N=100$, corresponding to physical volume $N/\sqrt{x} =100$ at lattice spacing $x=1$, for times after the meson-meson collision, $g t_{\mathrm{phys}}=100$. The initial momentum of the mesons, $k=1$, is below the momentum threshold. The fermion mass is $m/g=10^{-5}$, and the standard deviation of the initial wavepackets $\sigma=4$. We set $C^{mn}=0$ for $|m-n|\leq 1$, so that $L^{m,n}$ do not overlap. The insets show the enclosed area rescaled by a factor, to highlight the signal of the correlations between the two outgoing meson wavepackets of the elastic channel.
The bond dimension used for the calculations is $D=50$ and the maximum electric flux $L_{trunc}=8$. }
\label{energy correlator, below threshold}
\end{figure}

\begin{figure*}[t!]
\centering
\includegraphics[width=1.7\columnwidth]{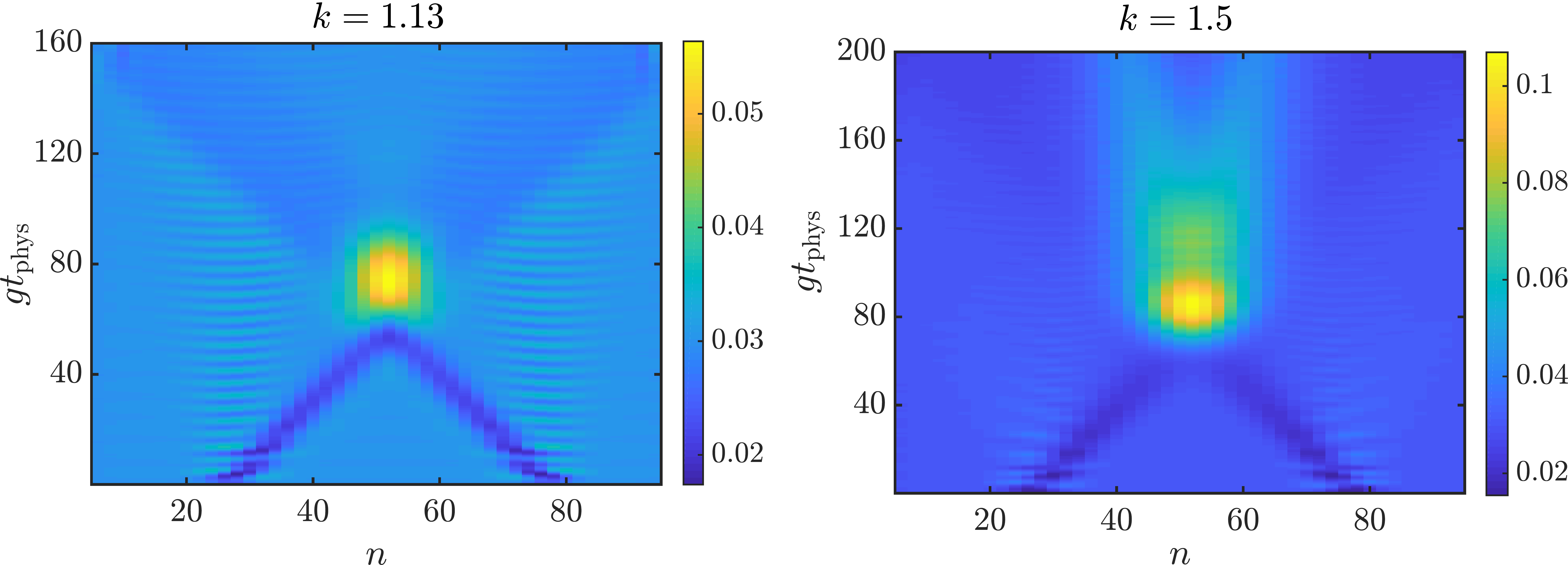}
\caption{This plot shows the expectation value of the four-body projector, $P_{n}(t)$, of a system with lattice size $N=100$ and physical volume $N /\sqrt{x} =100$. For the first plot the initial momenta of the mesons are $k=1.13$ and for the second plot are $k=1.5$. The mass is $m/g=10^{-5}$, and the standard deviation of the wavepackets $\sigma=4$. The bond dimension is $D=50$ and the maximum electric flux $L_{trunc}=8$. }
\label{projector}
\end{figure*}

\begin{figure}[h!]
\centering
\includegraphics[width=0.85\columnwidth]{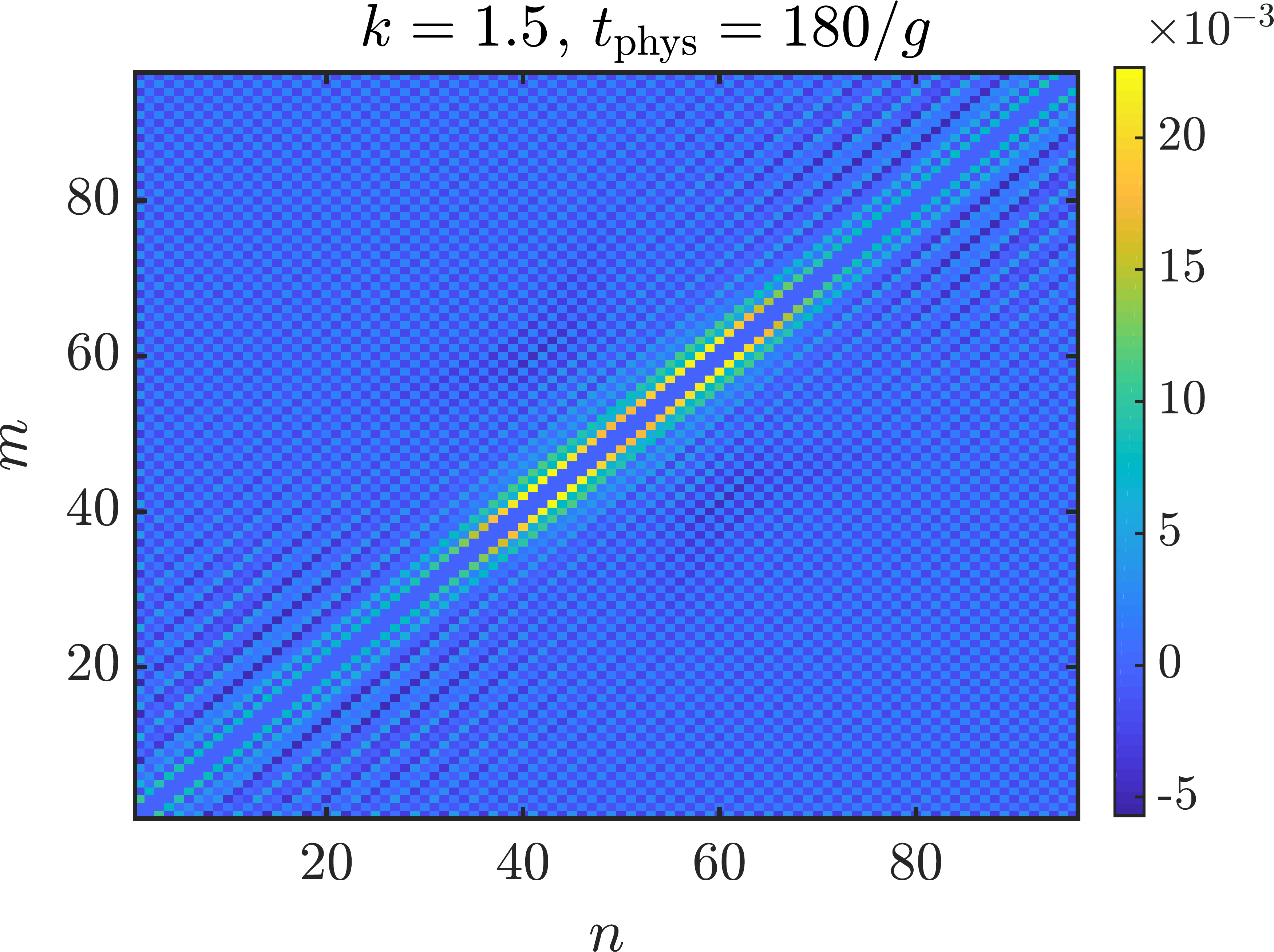}
\caption{This plot shows the electric flux correlator, $C^{mn}_{l}$, of a system size $N=100$ and physical volume $N /\sqrt{x} =100$, for times after the meson-meson collision, $gt_{\mathrm{phys}}=180$. The initial momenta of the mesons are above the momentum threshold, $k=1.5$. The mass is $m/g=10^{-5}$, and the standard deviation of the wavepackets $\sigma=4$. The bond dimension is $D=50$ and the maximum electric flux $L_{trunc}=8$. We set $C^{mn}=0$ for $m=n$, $m=n-1$ and $m=n+1$.}

\label{energy correlator, above threshold, 100}
\end{figure}

In this subsection we plot the electric flux correlator, 
\begin{align}
    \mathcal{C}^{mn}(t) &= \langle L^{m} L^{n} \rangle - \langle L^{m} \rangle \langle L^{n} \rangle
\end{align}
with $L^{n}=\frac{1}{2}\sum_{k=0}^{n}((-1)^{k}+\sigma_{k}^{z})$. 
In principle, we can expect that
this quantity can be measured in experiments, showing the signals of particle production.

The plots in this section show the electric flux correlators, $C^{mn}(t)$, specifically for times after the collision. Fig.~\ref{energy correlator, below threshold} shows $C^{mn}(t)$ for $g t_{\mathrm{phys}}=100$ for the case in which only the elastic channel appears, with the incoming mesons having momentum $k=1$. As seen in the left plot of Fig.~\ref{entanglement}, the outgoing mesons appear to be occupying the lattice sites $20-40$ and $60-80$. This is verified by Fig.~\ref{energy correlator, below threshold}, where the two oval shapes around the diagonal correspond to the two outgoing mesons of the elastic channel and are concentrated around the same positions.

In Fig.~\ref{energy correlator, above threshold, 100} we plot the correlator for $g t_{\mathrm{phys}}=180$ and initial meson momenta above the momentum threshold, at $k=1.5$.
In this case both the elastic and inelastic channel are present. The inelastic channel appears as the yellow signal around the main diagonal, with the two bound states visible around sites $40-50$ and $60-70$.

\subsection{Four-body projector}
\label{sec: bound states structure}
As seen in the subsections above, two different particles are produced when the momenta of the initial mesons are above the momentum threshold $k_{\mathrm{thr}}^{\mathrm{MPS}}\approx 1.12$. These are scalar mesons, which can be interpreted, as discussed in section~\ref{sec: Model, continuum} as bound states of two Schwinger bosons.

Ideally, to unambiguously identify the particles produced as scalar mesons, we would like to measure the corresponding scalar operator, which is not known, in general.
However, in the strong coupling regime, $m/g\rightarrow 0$, the continuum theory results ensure that the composition of the scalar meson in that case is dominated by four fermion terms, whereas the vector meson is dominated by two-fermion terms~\cite{mo1993basis,harada1995}. Even though we are far from that limit and both the ground states and excitations are strongly renormalised, we propose probing the appearance of the scalar mesons 
by calculating the time-dependent expectation value of the four-body projector:
\begin{align}
P_{n}(t)&= \langle \Pi_{\uparrow}^{n} \Pi_{\downarrow}^{n+1} \Pi_{\uparrow}^{n+2} \Pi_{\downarrow}^{n+3} \rangle |_{n=even}
\end{align}
with $\Pi_{\uparrow}^{n}$,  $\Pi_{\downarrow}^{n}$ the projectors:
\begin{align}
    \Pi_{\uparrow}^{n} |\uparrow \rangle = |\uparrow \rangle \\
    \Pi_{\downarrow}^{n} |\downarrow \rangle = |\downarrow \rangle \nonumber
\end{align}
and $\Pi_{\uparrow}^{n} |\downarrow \rangle =0$, $\Pi_{\downarrow}^{n} |\uparrow \rangle =0$. This will give a signal that the time-evolved state has a four-body component iff all the consecutive sites $n$, $n+1$, $n+2$ and $n+3$ are occupied. The underlying assumption is that the wave functions are continuously connected to the strong-coupling regime, such that we can identify its signal as the presence of the scalar.

The results shown in Fig.~\ref{projector} give yet another indication of the presence of the bound states. Specifically, in the left plot of Fig.~\ref{projector} ($k=1.13$) one observes the first signal of bound state production (yellow colour). This corresponds to two scalar mesons produced almost at rest (approximately zero velocities). The right plot of Fig.~\ref{projector} shows the case where the incoming mesons have momentum $k=1.5$. After the collision, one observes signals of two bound states with opposite velocities (yellow colour).

\section{Discussion}\label{sec: Discussion}

In this work we have used TNS methods to simulate scattering processes in the lattice version of the Schwinger model. More concretely, we have studied the scattering in the center of mass of two vector mesons (composite particles) at different incident momenta, and have been able to observe the threshold at which the inelastic channel opens and other particles of higher mass are produced. We have proposed entanglement and local observables to identify the character of the collision and the nature of the products. More concretely, we show that the entanglement of two neighboring sites (which would be accessible from measurements on two sites) with the rest of the chain and the correlators of the electric flux provide clear signatures of the collision character. Moreover, a 4-site projector can be used as another measure for observing the presence of the collision products.

Our initial state preparation and detection observables are based on perturbative expansions valid in the strong coupling (small fermion mass and small $x$ parameter) regime, and we performed our numerical simulations at a fixed value of the lattice spacing, far from the continuum limit. Yet, this parameter regime would be closer to the ones possible for current quantum simulation platforms, for example ultracold atoms as recently proposed for a related U(1) quantum link model~\cite{su2024cold}. 

Quantum inspired classical simulations with TNS, as well as proposed quantum simulation protocols are tackling progressively challenging regimes of LGT, and could in the future study problems that are out of the reach of traditional lattice methods. Out of equilibrium processes, such as collision experiments, are among the most interesting ones, but also come with their own challenges. In particular, performing a systematic extrapolation of lattice results to the continuum limit for a real-time problem, with a precision close to the one attained for spectral or thermal properties, requires performing real-time simulations on larger lattices far from the strong coupling limit as $x\to\infty$. It is thus a natural extension of this work to investigate how to prepare suitable initial states
and to identify other observables that can detect the products of the collision away from the strong coupling perturbative regime.

Further extensions of our work could involve studying the effect of a background electric field, which is known to induce a phase transition, or considering collisions at much higher energies, with the possibility to open other inelastic channels. Overall, we hope that our results provide a stepping stone for exploring real-time scattering phenomena in regimes inaccessible by perturbative methods. 

\begin{acknowledgments}
We thank Joseph Vovrosh, Krzysztof Cichy, Manuel Schneider and David C.-J. Lin for very helpful discussions. 
This work was partly supported by  the DFG (German Research Foundation) under Germany's Excellence Strategy -- EXC-2111 -- 390814868,  Research Unit FOR 5522 (grant nr. 499180199), and TRR 360 - 492547816;
and by the EU-QUANTERA project TNiSQ (BA 6059/1-1).
Research was supported by the Munich Quantum Valley, which is supported by the Bavarian state government with funds from the Hightech Agenda Bayern Plus. J.K. acknowledges support from the Imperial-TUM flagship partnership.
\end{acknowledgments}

\bibliography{refs.bib}

\begin{thebibliography}{67}%
\makeatletter
\providecommand \@ifxundefined [1]{%
 \@ifx{#1\undefined}
}%
\providecommand \@ifnum [1]{%
 \ifnum #1\expandafter \@firstoftwo
 \else \expandafter \@secondoftwo
 \fi
}%
\providecommand \@ifx [1]{%
 \ifx #1\expandafter \@firstoftwo
 \else \expandafter \@secondoftwo
 \fi
}%
\providecommand \natexlab [1]{#1}%
\providecommand \enquote  [1]{``#1''}%
\providecommand \bibnamefont  [1]{#1}%
\providecommand \bibfnamefont [1]{#1}%
\providecommand \citenamefont [1]{#1}%
\providecommand \href@noop [0]{\@secondoftwo}%
\providecommand \href [0]{\begingroup \@sanitize@url \@href}%
\providecommand \@href[1]{\@@startlink{#1}\@@href}%
\providecommand \@@href[1]{\endgroup#1\@@endlink}%
\providecommand \@sanitize@url [0]{\catcode `\\12\catcode `\$12\catcode
  `\&12\catcode `\#12\catcode `\^12\catcode `\_12\catcode `\%12\relax}%
\providecommand \@@startlink[1]{}%
\providecommand \@@endlink[0]{}%
\providecommand \url  [0]{\begingroup\@sanitize@url \@url }%
\providecommand \@url [1]{\endgroup\@href {#1}{\urlprefix }}%
\providecommand \urlprefix  [0]{URL }%
\providecommand \Eprint [0]{\href }%
\providecommand \doibase [0]{https://doi.org/}%
\providecommand \selectlanguage [0]{\@gobble}%
\providecommand \bibinfo  [0]{\@secondoftwo}%
\providecommand \bibfield  [0]{\@secondoftwo}%
\providecommand \translation [1]{[#1]}%
\providecommand \BibitemOpen [0]{}%
\providecommand \bibitemStop [0]{}%
\providecommand \bibitemNoStop [0]{.\EOS\space}%
\providecommand \EOS [0]{\spacefactor3000\relax}%
\providecommand \BibitemShut  [1]{\csname bibitem#1\endcsname}%
\let\auto@bib@innerbib\@empty
\bibitem [{\citenamefont {Brambilla}\ \emph {et~al.}(2014)\citenamefont
  {Brambilla}, \citenamefont {Eidelman}, \citenamefont {Foka}, \citenamefont
  {Gardner}, \citenamefont {Kronfeld}, \citenamefont {Alford}, \citenamefont
  {Alkofer}, \citenamefont {Butenschoen}, \citenamefont {Cohen}, \citenamefont
  {Erdmenger}, \citenamefont {Fabbietti}, \citenamefont {Faber}, \citenamefont
  {Goity}, \citenamefont {Ketzer}, \citenamefont {Lin}, \citenamefont
  {Llanes-Estrada}, \citenamefont {Meyer}, \citenamefont {Pakhlov},
  \citenamefont {Pallante}, \citenamefont {Polikarpov}, \citenamefont
  {Sazdjian}, \citenamefont {Schmitt}, \citenamefont {Snow}, \citenamefont
  {Vairo}, \citenamefont {Vogt}, \citenamefont {Vuorinen}, \citenamefont
  {Wittig}, \citenamefont {Arnold}, \citenamefont {Christakoglou},
  \citenamefont {Di~Nezza}, \citenamefont {Fodor}, \citenamefont {Garcia~i
  Tormo}, \citenamefont {H{\"o}llwieser}, \citenamefont {Janik}, \citenamefont
  {Kalweit}, \citenamefont {Keane}, \citenamefont {Kiritsis}, \citenamefont
  {Mischke}, \citenamefont {Mizuk}, \citenamefont {Odyniec}, \citenamefont
  {Papadodimas}, \citenamefont {Pich}, \citenamefont {Pittau}, \citenamefont
  {Qiu}, \citenamefont {Ricciardi}, \citenamefont {Salgado}, \citenamefont
  {Schwenzer}, \citenamefont {Stefanis}, \citenamefont {von Hippel},\ and\
  \citenamefont {Zakharov}}]{Brambilla2014}%
  \BibitemOpen
  \bibfield  {author} {\bibinfo {author} {\bibfnamefont {N.}~\bibnamefont
  {Brambilla}}, \bibinfo {author} {\bibfnamefont {S.}~\bibnamefont {Eidelman}},
  \bibinfo {author} {\bibfnamefont {P.}~\bibnamefont {Foka}}, \bibinfo {author}
  {\bibfnamefont {S.}~\bibnamefont {Gardner}}, \bibinfo {author} {\bibfnamefont
  {A.~S.}\ \bibnamefont {Kronfeld}}, \bibinfo {author} {\bibfnamefont {M.~G.}\
  \bibnamefont {Alford}}, \bibinfo {author} {\bibfnamefont {R.}~\bibnamefont
  {Alkofer}}, \bibinfo {author} {\bibfnamefont {M.}~\bibnamefont
  {Butenschoen}}, \bibinfo {author} {\bibfnamefont {T.~D.}\ \bibnamefont
  {Cohen}}, \bibinfo {author} {\bibfnamefont {J.}~\bibnamefont {Erdmenger}},
  \bibinfo {author} {\bibfnamefont {L.}~\bibnamefont {Fabbietti}}, \bibinfo
  {author} {\bibfnamefont {M.}~\bibnamefont {Faber}}, \bibinfo {author}
  {\bibfnamefont {J.~L.}\ \bibnamefont {Goity}}, \bibinfo {author}
  {\bibfnamefont {B.}~\bibnamefont {Ketzer}}, \bibinfo {author} {\bibfnamefont
  {H.~W.}\ \bibnamefont {Lin}}, \bibinfo {author} {\bibfnamefont {F.~J.}\
  \bibnamefont {Llanes-Estrada}}, \bibinfo {author} {\bibfnamefont {H.~B.}\
  \bibnamefont {Meyer}}, \bibinfo {author} {\bibfnamefont {P.}~\bibnamefont
  {Pakhlov}}, \bibinfo {author} {\bibfnamefont {E.}~\bibnamefont {Pallante}},
  \bibinfo {author} {\bibfnamefont {M.~I.}\ \bibnamefont {Polikarpov}},
  \bibinfo {author} {\bibfnamefont {H.}~\bibnamefont {Sazdjian}}, \bibinfo
  {author} {\bibfnamefont {A.}~\bibnamefont {Schmitt}}, \bibinfo {author}
  {\bibfnamefont {W.~M.}\ \bibnamefont {Snow}}, \bibinfo {author}
  {\bibfnamefont {A.}~\bibnamefont {Vairo}}, \bibinfo {author} {\bibfnamefont
  {R.}~\bibnamefont {Vogt}}, \bibinfo {author} {\bibfnamefont {A.}~\bibnamefont
  {Vuorinen}}, \bibinfo {author} {\bibfnamefont {H.}~\bibnamefont {Wittig}},
  \bibinfo {author} {\bibfnamefont {P.}~\bibnamefont {Arnold}}, \bibinfo
  {author} {\bibfnamefont {P.}~\bibnamefont {Christakoglou}}, \bibinfo {author}
  {\bibfnamefont {P.}~\bibnamefont {Di~Nezza}}, \bibinfo {author}
  {\bibfnamefont {Z.}~\bibnamefont {Fodor}}, \bibinfo {author} {\bibfnamefont
  {X.}~\bibnamefont {Garcia~i Tormo}}, \bibinfo {author} {\bibfnamefont
  {R.}~\bibnamefont {H{\"o}llwieser}}, \bibinfo {author} {\bibfnamefont
  {M.~A.}\ \bibnamefont {Janik}}, \bibinfo {author} {\bibfnamefont
  {A.}~\bibnamefont {Kalweit}}, \bibinfo {author} {\bibfnamefont
  {D.}~\bibnamefont {Keane}}, \bibinfo {author} {\bibfnamefont
  {E.}~\bibnamefont {Kiritsis}}, \bibinfo {author} {\bibfnamefont
  {A.}~\bibnamefont {Mischke}}, \bibinfo {author} {\bibfnamefont
  {R.}~\bibnamefont {Mizuk}}, \bibinfo {author} {\bibfnamefont
  {G.}~\bibnamefont {Odyniec}}, \bibinfo {author} {\bibfnamefont
  {K.}~\bibnamefont {Papadodimas}}, \bibinfo {author} {\bibfnamefont
  {A.}~\bibnamefont {Pich}}, \bibinfo {author} {\bibfnamefont {R.}~\bibnamefont
  {Pittau}}, \bibinfo {author} {\bibfnamefont {J.~W.}\ \bibnamefont {Qiu}},
  \bibinfo {author} {\bibfnamefont {G.}~\bibnamefont {Ricciardi}}, \bibinfo
  {author} {\bibfnamefont {C.~A.}\ \bibnamefont {Salgado}}, \bibinfo {author}
  {\bibfnamefont {K.}~\bibnamefont {Schwenzer}}, \bibinfo {author}
  {\bibfnamefont {N.~G.}\ \bibnamefont {Stefanis}}, \bibinfo {author}
  {\bibfnamefont {G.~M.}\ \bibnamefont {von Hippel}},\ and\ \bibinfo {author}
  {\bibfnamefont {V.~I.}\ \bibnamefont {Zakharov}},\ }\bibfield  {title}
  {\bibinfo {title} {{QCD and strongly coupled gauge theories: challenges and
  perspectives}},\ }\href {https://doi.org/10.1140/epjc/s10052-014-2981-5}
  {\bibfield  {journal} {\bibinfo  {journal} {The European Physical Journal C}\
  }\textbf {\bibinfo {volume} {74}},\ \bibinfo {pages} {2981} (\bibinfo {year}
  {2014})}\BibitemShut {NoStop}%
\bibitem [{\citenamefont {Wilson}(1974)}]{Wilson1974}%
  \BibitemOpen
  \bibfield  {author} {\bibinfo {author} {\bibfnamefont {K.~G.}\ \bibnamefont
  {Wilson}},\ }\bibfield  {title} {\bibinfo {title} {Confinement of quarks},\
  }\href {https://doi.org/10.1103/PhysRevD.10.2445} {\bibfield  {journal}
  {\bibinfo  {journal} {Phys. Rev. D}\ }\textbf {\bibinfo {volume} {10}},\
  \bibinfo {pages} {2445} (\bibinfo {year} {1974})}\BibitemShut {NoStop}%
\bibitem [{\citenamefont {Foulkes}\ \emph {et~al.}(2001)\citenamefont
  {Foulkes}, \citenamefont {Mitas}, \citenamefont {Needs},\ and\ \citenamefont
  {Rajagopal}}]{foulkes2001}%
  \BibitemOpen
  \bibfield  {author} {\bibinfo {author} {\bibfnamefont {W.~M.~C.}\
  \bibnamefont {Foulkes}}, \bibinfo {author} {\bibfnamefont {L.}~\bibnamefont
  {Mitas}}, \bibinfo {author} {\bibfnamefont {R.~J.}\ \bibnamefont {Needs}},\
  and\ \bibinfo {author} {\bibfnamefont {G.}~\bibnamefont {Rajagopal}},\
  }\bibfield  {title} {\bibinfo {title} {{Quantum Monte Carlo simulations of
  solids}},\ }\href {https://doi.org/10.1103/RevModPhys.73.33} {\bibfield
  {journal} {\bibinfo  {journal} {Rev. Mod. Phys.}\ }\textbf {\bibinfo {volume}
  {73}},\ \bibinfo {pages} {33} (\bibinfo {year} {2001})}\BibitemShut {NoStop}%
\bibitem [{\citenamefont {Carlson}\ \emph {et~al.}(2015)\citenamefont
  {Carlson}, \citenamefont {Gandolfi}, \citenamefont {Pederiva}, \citenamefont
  {Pieper}, \citenamefont {Schiavilla}, \citenamefont {Schmidt},\ and\
  \citenamefont {Wiringa}}]{carlson2015}%
  \BibitemOpen
  \bibfield  {author} {\bibinfo {author} {\bibfnamefont {J.}~\bibnamefont
  {Carlson}}, \bibinfo {author} {\bibfnamefont {S.}~\bibnamefont {Gandolfi}},
  \bibinfo {author} {\bibfnamefont {F.}~\bibnamefont {Pederiva}}, \bibinfo
  {author} {\bibfnamefont {S.~C.}\ \bibnamefont {Pieper}}, \bibinfo {author}
  {\bibfnamefont {R.}~\bibnamefont {Schiavilla}}, \bibinfo {author}
  {\bibfnamefont {K.~E.}\ \bibnamefont {Schmidt}},\ and\ \bibinfo {author}
  {\bibfnamefont {R.~B.}\ \bibnamefont {Wiringa}},\ }\bibfield  {title}
  {\bibinfo {title} {{Quantum Monte Carlo methods for nuclear physics}},\
  }\href {https://doi.org/10.1103/RevModPhys.87.1067} {\bibfield  {journal}
  {\bibinfo  {journal} {Rev. Mod. Phys.}\ }\textbf {\bibinfo {volume} {87}},\
  \bibinfo {pages} {1067} (\bibinfo {year} {2015})}\BibitemShut {NoStop}%
\bibitem [{\citenamefont {Cichy}\ and\ \citenamefont
  {Constantinou}(2019)}]{Cichy2019}%
  \BibitemOpen
  \bibfield  {author} {\bibinfo {author} {\bibfnamefont {K.}~\bibnamefont
  {Cichy}}\ and\ \bibinfo {author} {\bibfnamefont {M.}~\bibnamefont
  {Constantinou}},\ }\bibfield  {title} {\bibinfo {title} {{A Guide to
  Light-Cone PDFs from Lattice QCD: An Overview of Approaches, Techniques, and
  Results}},\ }\href {https://doi.org/10.1155/2019/3036904} {\bibfield
  {journal} {\bibinfo  {journal} {Advances in High Energy Physics}\ }\textbf
  {\bibinfo {volume} {2019}},\ \bibinfo {pages} {3036904} (\bibinfo {year}
  {2019})}\BibitemShut {NoStop}%
\bibitem [{\citenamefont {Ba{\~{n}}uls}\ and\ \citenamefont
  {Cichy}(2020)}]{banuls2020ropp}%
  \BibitemOpen
  \bibfield  {author} {\bibinfo {author} {\bibfnamefont {M.~C.}\ \bibnamefont
  {Ba{\~{n}}uls}}\ and\ \bibinfo {author} {\bibfnamefont {K.}~\bibnamefont
  {Cichy}},\ }\bibfield  {title} {\bibinfo {title} {Review on novel methods for
  lattice gauge theories},\ }\href
  {https://doi.org/https://doi.org/10.1088/1361-6633/ab6311} {\bibfield
  {journal} {\bibinfo  {journal} {Reports on Progress in Physics}\ }\textbf
  {\bibinfo {volume} {83}},\ \bibinfo {pages} {024401} (\bibinfo {year}
  {2020})}\BibitemShut {NoStop}%
\bibitem [{\citenamefont {{Ba\~nuls, Mari Carmen}}\ \emph
  {et~al.}(2020)\citenamefont {{Ba\~nuls, Mari Carmen}}, \citenamefont {{Blatt,
  Rainer}}, \citenamefont {{Catani, Jacopo}}, \citenamefont {{Celi, Alessio}},
  \citenamefont {{Cirac, Juan Ignacio}}, \citenamefont {{Dalmonte, Marcello}},
  \citenamefont {{Fallani, Leonardo}}, \citenamefont {{Jansen, Karl}},
  \citenamefont {{Lewenstein, Maciej}}, \citenamefont {{Montangero, Simone}},
  \citenamefont {{Muschik, Christine A.}}, \citenamefont {{Reznik, Benni}},
  \citenamefont {{Rico, Enrique}}, \citenamefont {{Tagliacozzo, Luca}},
  \citenamefont {{Van Acoleyen, Karel}}, \citenamefont {{Verstraete, Frank}},
  \citenamefont {{Wiese, Uwe-Jens}}, \citenamefont {{Wingate, Matthew}},
  \citenamefont {{Zakrzewski, Jakub}},\ and\ \citenamefont {{Zoller,
  Peter}}}]{banuls2020epj}%
  \BibitemOpen
  \bibfield  {author} {\bibinfo {author} {\bibnamefont {{Ba\~nuls, Mari
  Carmen}}}, \bibinfo {author} {\bibnamefont {{Blatt, Rainer}}}, \bibinfo
  {author} {\bibnamefont {{Catani, Jacopo}}}, \bibinfo {author} {\bibnamefont
  {{Celi, Alessio}}}, \bibinfo {author} {\bibnamefont {{Cirac, Juan Ignacio}}},
  \bibinfo {author} {\bibnamefont {{Dalmonte, Marcello}}}, \bibinfo {author}
  {\bibnamefont {{Fallani, Leonardo}}}, \bibinfo {author} {\bibnamefont
  {{Jansen, Karl}}}, \bibinfo {author} {\bibnamefont {{Lewenstein, Maciej}}},
  \bibinfo {author} {\bibnamefont {{Montangero, Simone}}}, \bibinfo {author}
  {\bibnamefont {{Muschik, Christine A.}}}, \bibinfo {author} {\bibnamefont
  {{Reznik, Benni}}}, \bibinfo {author} {\bibnamefont {{Rico, Enrique}}},
  \bibinfo {author} {\bibnamefont {{Tagliacozzo, Luca}}}, \bibinfo {author}
  {\bibnamefont {{Van Acoleyen, Karel}}}, \bibinfo {author} {\bibnamefont
  {{Verstraete, Frank}}}, \bibinfo {author} {\bibnamefont {{Wiese, Uwe-Jens}}},
  \bibinfo {author} {\bibnamefont {{Wingate, Matthew}}}, \bibinfo {author}
  {\bibnamefont {{Zakrzewski, Jakub}}},\ and\ \bibinfo {author} {\bibnamefont
  {{Zoller, Peter}}},\ }\bibfield  {title} {\bibinfo {title} {Simulating
  lattice gauge theories within quantum technologies},\ }\href
  {https://doi.org/https://doi.org/10.1140/epjd/e2020-100571-8} {\bibfield
  {journal} {\bibinfo  {journal} {Eur. Phys. J. D}\ }\textbf {\bibinfo {volume}
  {74}},\ \bibinfo {pages} {165} (\bibinfo {year} {2020})}\BibitemShut
  {NoStop}%
\bibitem [{\citenamefont {Davoudi}\ \emph {et~al.}(2020)\citenamefont
  {Davoudi}, \citenamefont {Hafezi}, \citenamefont {Monroe}, \citenamefont
  {Pagano}, \citenamefont {Seif},\ and\ \citenamefont {Shaw}}]{davoudi2020}%
  \BibitemOpen
  \bibfield  {author} {\bibinfo {author} {\bibfnamefont {Z.}~\bibnamefont
  {Davoudi}}, \bibinfo {author} {\bibfnamefont {M.}~\bibnamefont {Hafezi}},
  \bibinfo {author} {\bibfnamefont {C.}~\bibnamefont {Monroe}}, \bibinfo
  {author} {\bibfnamefont {G.}~\bibnamefont {Pagano}}, \bibinfo {author}
  {\bibfnamefont {A.}~\bibnamefont {Seif}},\ and\ \bibinfo {author}
  {\bibfnamefont {A.}~\bibnamefont {Shaw}},\ }\bibfield  {title} {\bibinfo
  {title} {Towards analog quantum simulations of lattice gauge theories with
  trapped ions},\ }\href {https://doi.org/10.1103/PhysRevResearch.2.023015}
  {\bibfield  {journal} {\bibinfo  {journal} {Phys. Rev. Res.}\ }\textbf
  {\bibinfo {volume} {2}},\ \bibinfo {pages} {023015} (\bibinfo {year}
  {2020})}\BibitemShut {NoStop}%
\bibitem [{\citenamefont {Bass}\ and\ \citenamefont
  {Zohar}(2022)}]{bass2021qtech}%
  \BibitemOpen
  \bibfield  {author} {\bibinfo {author} {\bibfnamefont {S.~D.}\ \bibnamefont
  {Bass}}\ and\ \bibinfo {author} {\bibfnamefont {E.}~\bibnamefont {Zohar}},\
  }\bibfield  {title} {\bibinfo {title} {Quantum technologies in particle
  physics},\ }\href {https://doi.org/https://doi.org/10.1098/rsta.2021.0072}
  {\bibfield  {journal} {\bibinfo  {journal} {Phil. Trans. R. Soc. A}\ }\textbf
  {\bibinfo {volume} {380}},\ \bibinfo {pages} {20210072} (\bibinfo {year}
  {2022})}\BibitemShut {NoStop}%
\bibitem [{\citenamefont {Funcke}\ \emph {et~al.}(2023)\citenamefont {Funcke},
  \citenamefont {Hartung}, \citenamefont {Jansen},\ and\ \citenamefont
  {Kühn}}]{Funcke2023}%
  \BibitemOpen
  \bibfield  {author} {\bibinfo {author} {\bibfnamefont {L.}~\bibnamefont
  {Funcke}}, \bibinfo {author} {\bibfnamefont {T.}~\bibnamefont {Hartung}},
  \bibinfo {author} {\bibfnamefont {K.}~\bibnamefont {Jansen}},\ and\ \bibinfo
  {author} {\bibfnamefont {S.}~\bibnamefont {Kühn}},\ }\bibfield  {title}
  {\bibinfo {title} {{{Review on Quantum Computing for Lattice Field
  Theory}}},\ }\href {https://doi.org/10.22323/1.430.0228} {\bibfield
  {journal} {\bibinfo  {journal} {PoS}\ }\textbf {\bibinfo {volume}
  {LATTICE2022}},\ \bibinfo {pages} {228} (\bibinfo {year} {2023})}\BibitemShut
  {NoStop}%
\bibitem [{\citenamefont {Bauer}\ \emph {et~al.}(2023)\citenamefont {Bauer},
  \citenamefont {Davoudi}, \citenamefont {Balantekin}, \citenamefont
  {Bhattacharya}, \citenamefont {Carena}, \citenamefont {de~Jong},
  \citenamefont {Draper}, \citenamefont {El-Khadra}, \citenamefont {Gemelke},
  \citenamefont {Hanada}, \citenamefont {Kharzeev}, \citenamefont {Lamm},
  \citenamefont {Li}, \citenamefont {Liu}, \citenamefont {Lukin}, \citenamefont
  {Meurice}, \citenamefont {Monroe}, \citenamefont {Nachman}, \citenamefont
  {Pagano}, \citenamefont {Preskill}, \citenamefont {Rinaldi}, \citenamefont
  {Roggero}, \citenamefont {Santiago}, \citenamefont {Savage}, \citenamefont
  {Siddiqi}, \citenamefont {Siopsis}, \citenamefont {Van~Zanten}, \citenamefont
  {Wiebe}, \citenamefont {Yamauchi}, \citenamefont {Yeter-Aydeniz},\ and\
  \citenamefont {Zorzetti}}]{Bauer2023}%
  \BibitemOpen
  \bibfield  {author} {\bibinfo {author} {\bibfnamefont {C.~W.}\ \bibnamefont
  {Bauer}}, \bibinfo {author} {\bibfnamefont {Z.}~\bibnamefont {Davoudi}},
  \bibinfo {author} {\bibfnamefont {A.~B.}\ \bibnamefont {Balantekin}},
  \bibinfo {author} {\bibfnamefont {T.}~\bibnamefont {Bhattacharya}}, \bibinfo
  {author} {\bibfnamefont {M.}~\bibnamefont {Carena}}, \bibinfo {author}
  {\bibfnamefont {W.~A.}\ \bibnamefont {de~Jong}}, \bibinfo {author}
  {\bibfnamefont {P.}~\bibnamefont {Draper}}, \bibinfo {author} {\bibfnamefont
  {A.}~\bibnamefont {El-Khadra}}, \bibinfo {author} {\bibfnamefont
  {N.}~\bibnamefont {Gemelke}}, \bibinfo {author} {\bibfnamefont
  {M.}~\bibnamefont {Hanada}}, \bibinfo {author} {\bibfnamefont
  {D.}~\bibnamefont {Kharzeev}}, \bibinfo {author} {\bibfnamefont
  {H.}~\bibnamefont {Lamm}}, \bibinfo {author} {\bibfnamefont {Y.-Y.}\
  \bibnamefont {Li}}, \bibinfo {author} {\bibfnamefont {J.}~\bibnamefont
  {Liu}}, \bibinfo {author} {\bibfnamefont {M.}~\bibnamefont {Lukin}}, \bibinfo
  {author} {\bibfnamefont {Y.}~\bibnamefont {Meurice}}, \bibinfo {author}
  {\bibfnamefont {C.}~\bibnamefont {Monroe}}, \bibinfo {author} {\bibfnamefont
  {B.}~\bibnamefont {Nachman}}, \bibinfo {author} {\bibfnamefont
  {G.}~\bibnamefont {Pagano}}, \bibinfo {author} {\bibfnamefont
  {J.}~\bibnamefont {Preskill}}, \bibinfo {author} {\bibfnamefont
  {E.}~\bibnamefont {Rinaldi}}, \bibinfo {author} {\bibfnamefont
  {A.}~\bibnamefont {Roggero}}, \bibinfo {author} {\bibfnamefont {D.~I.}\
  \bibnamefont {Santiago}}, \bibinfo {author} {\bibfnamefont {M.~J.}\
  \bibnamefont {Savage}}, \bibinfo {author} {\bibfnamefont {I.}~\bibnamefont
  {Siddiqi}}, \bibinfo {author} {\bibfnamefont {G.}~\bibnamefont {Siopsis}},
  \bibinfo {author} {\bibfnamefont {D.}~\bibnamefont {Van~Zanten}}, \bibinfo
  {author} {\bibfnamefont {N.}~\bibnamefont {Wiebe}}, \bibinfo {author}
  {\bibfnamefont {Y.}~\bibnamefont {Yamauchi}}, \bibinfo {author}
  {\bibfnamefont {K.}~\bibnamefont {Yeter-Aydeniz}},\ and\ \bibinfo {author}
  {\bibfnamefont {S.}~\bibnamefont {Zorzetti}},\ }\bibfield  {title} {\bibinfo
  {title} {{Quantum Simulation for High-Energy Physics}},\ }\href
  {https://doi.org/10.1103/PRXQuantum.4.027001} {\bibfield  {journal} {\bibinfo
   {journal} {PRX Quantum}\ }\textbf {\bibinfo {volume} {4}},\ \bibinfo {pages}
  {027001} (\bibinfo {year} {2023})}\BibitemShut {NoStop}%
\bibitem [{\citenamefont {Halimeh}\ \emph {et~al.}(2023)\citenamefont
  {Halimeh}, \citenamefont {Aidelsburger}, \citenamefont {Grusdt},
  \citenamefont {Hauke},\ and\ \citenamefont {Yang}}]{halimeh2023coldatom}%
  \BibitemOpen
  \bibfield  {author} {\bibinfo {author} {\bibfnamefont {J.~C.}\ \bibnamefont
  {Halimeh}}, \bibinfo {author} {\bibfnamefont {M.}~\bibnamefont
  {Aidelsburger}}, \bibinfo {author} {\bibfnamefont {F.}~\bibnamefont
  {Grusdt}}, \bibinfo {author} {\bibfnamefont {P.}~\bibnamefont {Hauke}},\ and\
  \bibinfo {author} {\bibfnamefont {B.}~\bibnamefont {Yang}},\ }\href@noop {}
  {\bibinfo {title} {{Cold-atom quantum simulators of gauge theories}}}
  (\bibinfo {year} {2023}),\ \Eprint {https://arxiv.org/abs/2310.12201}
  {arXiv:2310.12201} \BibitemShut {NoStop}%
\bibitem [{\citenamefont {Meglio}\ \emph {et~al.}(2023)\citenamefont {Meglio},
  \citenamefont {Jansen}, \citenamefont {Tavernelli}, \citenamefont
  {Alexandrou}, \citenamefont {Arunachalam}, \citenamefont {Bauer},
  \citenamefont {Borras}, \citenamefont {Carrazza}, \citenamefont {Crippa},
  \citenamefont {Croft}, \citenamefont {de~Putter}, \citenamefont {Delgado},
  \citenamefont {Dunjko}, \citenamefont {Egger}, \citenamefont
  {Fernandez-Combarro}, \citenamefont {Fuchs}, \citenamefont {Funcke},
  \citenamefont {Gonzalez-Cuadra}, \citenamefont {Grossi}, \citenamefont
  {Halimeh}, \citenamefont {Holmes}, \citenamefont {Kuhn}, \citenamefont
  {Lacroix}, \citenamefont {Lewis}, \citenamefont {Lucchesi}, \citenamefont
  {Martinez}, \citenamefont {Meloni}, \citenamefont {Mezzacapo}, \citenamefont
  {Montangero}, \citenamefont {Nagano}, \citenamefont {Radescu}, \citenamefont
  {Ortega}, \citenamefont {Roggero}, \citenamefont {Schuhmacher}, \citenamefont
  {Seixas}, \citenamefont {Silvi}, \citenamefont {Spentzouris}, \citenamefont
  {Tacchino}, \citenamefont {Temme}, \citenamefont {Terashi}, \citenamefont
  {Tura}, \citenamefont {Tuysuz}, \citenamefont {Vallecorsa}, \citenamefont
  {Wiese}, \citenamefont {Yoo},\ and\ \citenamefont
  {Zhang}}]{dimeglio2023quantum}%
  \BibitemOpen
  \bibfield  {author} {\bibinfo {author} {\bibfnamefont {A.~D.}\ \bibnamefont
  {Meglio}}, \bibinfo {author} {\bibfnamefont {K.}~\bibnamefont {Jansen}},
  \bibinfo {author} {\bibfnamefont {I.}~\bibnamefont {Tavernelli}}, \bibinfo
  {author} {\bibfnamefont {C.}~\bibnamefont {Alexandrou}}, \bibinfo {author}
  {\bibfnamefont {S.}~\bibnamefont {Arunachalam}}, \bibinfo {author}
  {\bibfnamefont {C.~W.}\ \bibnamefont {Bauer}}, \bibinfo {author}
  {\bibfnamefont {K.}~\bibnamefont {Borras}}, \bibinfo {author} {\bibfnamefont
  {S.}~\bibnamefont {Carrazza}}, \bibinfo {author} {\bibfnamefont
  {A.}~\bibnamefont {Crippa}}, \bibinfo {author} {\bibfnamefont
  {V.}~\bibnamefont {Croft}}, \bibinfo {author} {\bibfnamefont
  {R.}~\bibnamefont {de~Putter}}, \bibinfo {author} {\bibfnamefont
  {A.}~\bibnamefont {Delgado}}, \bibinfo {author} {\bibfnamefont
  {V.}~\bibnamefont {Dunjko}}, \bibinfo {author} {\bibfnamefont {D.~J.}\
  \bibnamefont {Egger}}, \bibinfo {author} {\bibfnamefont {E.}~\bibnamefont
  {Fernandez-Combarro}}, \bibinfo {author} {\bibfnamefont {E.}~\bibnamefont
  {Fuchs}}, \bibinfo {author} {\bibfnamefont {L.}~\bibnamefont {Funcke}},
  \bibinfo {author} {\bibfnamefont {D.}~\bibnamefont {Gonzalez-Cuadra}},
  \bibinfo {author} {\bibfnamefont {M.}~\bibnamefont {Grossi}}, \bibinfo
  {author} {\bibfnamefont {J.~C.}\ \bibnamefont {Halimeh}}, \bibinfo {author}
  {\bibfnamefont {Z.}~\bibnamefont {Holmes}}, \bibinfo {author} {\bibfnamefont
  {S.}~\bibnamefont {Kuhn}}, \bibinfo {author} {\bibfnamefont {D.}~\bibnamefont
  {Lacroix}}, \bibinfo {author} {\bibfnamefont {R.}~\bibnamefont {Lewis}},
  \bibinfo {author} {\bibfnamefont {D.}~\bibnamefont {Lucchesi}}, \bibinfo
  {author} {\bibfnamefont {M.~L.}\ \bibnamefont {Martinez}}, \bibinfo {author}
  {\bibfnamefont {F.}~\bibnamefont {Meloni}}, \bibinfo {author} {\bibfnamefont
  {A.}~\bibnamefont {Mezzacapo}}, \bibinfo {author} {\bibfnamefont
  {S.}~\bibnamefont {Montangero}}, \bibinfo {author} {\bibfnamefont
  {L.}~\bibnamefont {Nagano}}, \bibinfo {author} {\bibfnamefont
  {V.}~\bibnamefont {Radescu}}, \bibinfo {author} {\bibfnamefont {E.~R.}\
  \bibnamefont {Ortega}}, \bibinfo {author} {\bibfnamefont {A.}~\bibnamefont
  {Roggero}}, \bibinfo {author} {\bibfnamefont {J.}~\bibnamefont
  {Schuhmacher}}, \bibinfo {author} {\bibfnamefont {J.}~\bibnamefont {Seixas}},
  \bibinfo {author} {\bibfnamefont {P.}~\bibnamefont {Silvi}}, \bibinfo
  {author} {\bibfnamefont {P.}~\bibnamefont {Spentzouris}}, \bibinfo {author}
  {\bibfnamefont {F.}~\bibnamefont {Tacchino}}, \bibinfo {author}
  {\bibfnamefont {K.}~\bibnamefont {Temme}}, \bibinfo {author} {\bibfnamefont
  {K.}~\bibnamefont {Terashi}}, \bibinfo {author} {\bibfnamefont
  {J.}~\bibnamefont {Tura}}, \bibinfo {author} {\bibfnamefont {C.}~\bibnamefont
  {Tuysuz}}, \bibinfo {author} {\bibfnamefont {S.}~\bibnamefont {Vallecorsa}},
  \bibinfo {author} {\bibfnamefont {U.-J.}\ \bibnamefont {Wiese}}, \bibinfo
  {author} {\bibfnamefont {S.}~\bibnamefont {Yoo}},\ and\ \bibinfo {author}
  {\bibfnamefont {J.}~\bibnamefont {Zhang}},\ }\href@noop {} {\bibinfo {title}
  {{Quantum Computing for High-Energy Physics: State of the Art and Challenges.
  Summary of the QC4HEP Working Group}}} (\bibinfo {year} {2023}),\ \Eprint
  {https://arxiv.org/abs/2307.03236} {arXiv:2307.03236} \BibitemShut {NoStop}%
\bibitem [{\citenamefont {Kogut}(1979)}]{kogut1979introduction}%
  \BibitemOpen
  \bibfield  {author} {\bibinfo {author} {\bibfnamefont {J.~B.}\ \bibnamefont
  {Kogut}},\ }\bibfield  {title} {\bibinfo {title} {An introduction to lattice
  gauge theory and spin systems},\ }\href
  {https://doi.org/10.1103/RevModPhys.51.659} {\bibfield  {journal} {\bibinfo
  {journal} {Reviews of Modern Physics}\ }\textbf {\bibinfo {volume} {51}},\
  \bibinfo {pages} {659} (\bibinfo {year} {1979})}\BibitemShut {NoStop}%
\bibitem [{\citenamefont {Verstraete}\ \emph {et~al.}(2008)\citenamefont
  {Verstraete}, \citenamefont {Murg},\ and\ \citenamefont
  {Cirac}}]{Verstraete2008}%
  \BibitemOpen
  \bibfield  {author} {\bibinfo {author} {\bibfnamefont {F.}~\bibnamefont
  {Verstraete}}, \bibinfo {author} {\bibfnamefont {V.}~\bibnamefont {Murg}},\
  and\ \bibinfo {author} {\bibfnamefont {J.}~\bibnamefont {Cirac}},\ }\bibfield
   {title} {\bibinfo {title} {Matrix product states, projected entangled pair
  states, and variational renormalization group methods for quantum spin
  systems},\ }\href {https://doi.org/https://doi.org/10.1080/14789940801912366}
  {\bibfield  {journal} {\bibinfo  {journal} {Adv. Phys}\ }\textbf {\bibinfo
  {volume} {57}},\ \bibinfo {pages} {143} (\bibinfo {year} {2008})}\BibitemShut
  {NoStop}%
\bibitem [{\citenamefont {Schollwöck}(2011)}]{SCHOLLWOCK201196}%
  \BibitemOpen
  \bibfield  {author} {\bibinfo {author} {\bibfnamefont {U.}~\bibnamefont
  {Schollwöck}},\ }\bibfield  {title} {\bibinfo {title} {{The density-matrix
  renormalization group in the age of matrix product states}},\ }\href
  {https://doi.org/https://doi.org/10.1016/j.aop.2010.09.012} {\bibfield
  {journal} {\bibinfo  {journal} {Annals of Physics}\ }\textbf {\bibinfo
  {volume} {326}},\ \bibinfo {pages} {96} (\bibinfo {year} {2011})},\ \bibinfo
  {note} {january 2011 Special Issue}\BibitemShut {NoStop}%
\bibitem [{\citenamefont {Or{\'u}s}(2014)}]{Orus2014annphys}%
  \BibitemOpen
  \bibfield  {author} {\bibinfo {author} {\bibfnamefont {R.}~\bibnamefont
  {Or{\'u}s}},\ }\bibfield  {title} {\bibinfo {title} {{A practical
  introduction to tensor networks: Matrix product states and projected
  entangled pair states}},\ }\href
  {https://doi.org/https://doi.org/10.1016/j.aop.2014.06.013} {\bibfield
  {journal} {\bibinfo  {journal} {Annals Phys.}\ }\textbf {\bibinfo {volume}
  {349}},\ \bibinfo {pages} {117} (\bibinfo {year} {2014})}\BibitemShut
  {NoStop}%
\bibitem [{\citenamefont {Silvi}\ \emph {et~al.}(2019)\citenamefont {Silvi},
  \citenamefont {Tschirsich}, \citenamefont {Gerster}, \citenamefont
  {J{\"u}nemann}, \citenamefont {Jaschke}, \citenamefont {Rizzi},\ and\
  \citenamefont {Montangero}}]{Silvi2019tn}%
  \BibitemOpen
  \bibfield  {author} {\bibinfo {author} {\bibfnamefont {P.}~\bibnamefont
  {Silvi}}, \bibinfo {author} {\bibfnamefont {F.}~\bibnamefont {Tschirsich}},
  \bibinfo {author} {\bibfnamefont {M.}~\bibnamefont {Gerster}}, \bibinfo
  {author} {\bibfnamefont {J.}~\bibnamefont {J{\"u}nemann}}, \bibinfo {author}
  {\bibfnamefont {D.}~\bibnamefont {Jaschke}}, \bibinfo {author} {\bibfnamefont
  {M.}~\bibnamefont {Rizzi}},\ and\ \bibinfo {author} {\bibfnamefont
  {S.}~\bibnamefont {Montangero}},\ }\bibfield  {title} {\bibinfo {title} {{The
  Tensor Networks Anthology: Simulation techniques for many-body quantum
  lattice systems}},\ }\href
  {https://doi.org/https://doi.org/10.21468/SciPostPhysLectNotes.8} {\bibfield
  {journal} {\bibinfo  {journal} {SciPost Phys. Lect. Notes}\ ,\ \bibinfo
  {pages} {8}} (\bibinfo {year} {2019})}\BibitemShut {NoStop}%
\bibitem [{\citenamefont {Okunishi}\ \emph {et~al.}(2022)\citenamefont
  {Okunishi}, \citenamefont {Nishino},\ and\ \citenamefont
  {Ueda}}]{Okunishi2022}%
  \BibitemOpen
  \bibfield  {author} {\bibinfo {author} {\bibfnamefont {K.}~\bibnamefont
  {Okunishi}}, \bibinfo {author} {\bibfnamefont {T.}~\bibnamefont {Nishino}},\
  and\ \bibinfo {author} {\bibfnamefont {H.}~\bibnamefont {Ueda}},\ }\bibfield
  {title} {\bibinfo {title} {{Developments in the Tensor Network --- from
  Statistical Mechanics to Quantum Entanglement}},\ }\href
  {https://doi.org/https://doi.org/10.7566/JPSJ.91.062001} {\bibfield
  {journal} {\bibinfo  {journal} {J. Phys. Soc. Jpn}\ }\textbf {\bibinfo
  {volume} {91}},\ \bibinfo {pages} {062001} (\bibinfo {year}
  {2022})}\BibitemShut {NoStop}%
\bibitem [{\citenamefont {Ba\~{n}uls}(2023)}]{Banuls2023}%
  \BibitemOpen
  \bibfield  {author} {\bibinfo {author} {\bibfnamefont {M.~C.}\ \bibnamefont
  {Ba\~{n}uls}},\ }\bibfield  {title} {\bibinfo {title} {{Tensor Network
  Algorithms: A Route Map}},\ }\href@noop {} {\bibfield  {journal} {\bibinfo
  {journal} {Annu. Rev. Condens. Matter Phys}\ }\textbf {\bibinfo {volume}
  {14}} (\bibinfo {year} {2023})}\BibitemShut {NoStop}%
\bibitem [{\citenamefont {Felser}\ \emph {et~al.}(2020)\citenamefont {Felser},
  \citenamefont {Silvi}, \citenamefont {Collura},\ and\ \citenamefont
  {Montangero}}]{felser2020u1}%
  \BibitemOpen
  \bibfield  {author} {\bibinfo {author} {\bibfnamefont {T.}~\bibnamefont
  {Felser}}, \bibinfo {author} {\bibfnamefont {P.}~\bibnamefont {Silvi}},
  \bibinfo {author} {\bibfnamefont {M.}~\bibnamefont {Collura}},\ and\ \bibinfo
  {author} {\bibfnamefont {S.}~\bibnamefont {Montangero}},\ }\bibfield  {title}
  {\bibinfo {title} {{Two-Dimensional Quantum-Link Lattice Quantum
  Electrodynamics at Finite Density}},\ }\href
  {https://doi.org/10.1103/PhysRevX.10.041040} {\bibfield  {journal} {\bibinfo
  {journal} {Phys. Rev. X}\ }\textbf {\bibinfo {volume} {10}},\ \bibinfo
  {pages} {041040} (\bibinfo {year} {2020})}\BibitemShut {NoStop}%
\bibitem [{\citenamefont {Emonts}\ \emph {et~al.}(2020)\citenamefont {Emonts},
  \citenamefont {Ba\~nuls}, \citenamefont {Cirac},\ and\ \citenamefont
  {Zohar}}]{emonts2020z3}%
  \BibitemOpen
  \bibfield  {author} {\bibinfo {author} {\bibfnamefont {P.}~\bibnamefont
  {Emonts}}, \bibinfo {author} {\bibfnamefont {M.~C.}\ \bibnamefont
  {Ba\~nuls}}, \bibinfo {author} {\bibfnamefont {I.}~\bibnamefont {Cirac}},\
  and\ \bibinfo {author} {\bibfnamefont {E.}~\bibnamefont {Zohar}},\ }\bibfield
   {title} {\bibinfo {title} {{Variational Monte Carlo simulation with tensor
  networks of a pure ${\mathbb{Z}}_{3}$ gauge theory in $(2+1)\mathrm{D}$}},\
  }\href {https://doi.org/10.1103/PhysRevD.102.074501} {\bibfield  {journal}
  {\bibinfo  {journal} {Phys. Rev. D}\ }\textbf {\bibinfo {volume} {102}},\
  \bibinfo {pages} {074501} (\bibinfo {year} {2020})}\BibitemShut {NoStop}%
\bibitem [{\citenamefont {Robaina}\ \emph {et~al.}(2021)\citenamefont
  {Robaina}, \citenamefont {Ba\~nuls},\ and\ \citenamefont
  {Cirac}}]{robaina2021z3}%
  \BibitemOpen
  \bibfield  {author} {\bibinfo {author} {\bibfnamefont {D.}~\bibnamefont
  {Robaina}}, \bibinfo {author} {\bibfnamefont {M.~C.}\ \bibnamefont
  {Ba\~nuls}},\ and\ \bibinfo {author} {\bibfnamefont {J.~I.}\ \bibnamefont
  {Cirac}},\ }\bibfield  {title} {\bibinfo {title} {{Simulating $2+1\mathrm{D}$
  ${Z}_{3}$ Lattice Gauge Theory with an Infinite Projected Entangled-Pair
  State}},\ }\href {https://doi.org/10.1103/PhysRevLett.126.050401} {\bibfield
  {journal} {\bibinfo  {journal} {Phys. Rev. Lett.}\ }\textbf {\bibinfo
  {volume} {126}},\ \bibinfo {pages} {050401} (\bibinfo {year}
  {2021})}\BibitemShut {NoStop}%
\bibitem [{\citenamefont {Magnifico}\ \emph {et~al.}(2021)\citenamefont
  {Magnifico}, \citenamefont {Felser}, \citenamefont {Silvi},\ and\
  \citenamefont {Montangero}}]{magnifico2021qed3d}%
  \BibitemOpen
  \bibfield  {author} {\bibinfo {author} {\bibfnamefont {G.}~\bibnamefont
  {Magnifico}}, \bibinfo {author} {\bibfnamefont {T.}~\bibnamefont {Felser}},
  \bibinfo {author} {\bibfnamefont {P.}~\bibnamefont {Silvi}},\ and\ \bibinfo
  {author} {\bibfnamefont {S.}~\bibnamefont {Montangero}},\ }\bibfield  {title}
  {\bibinfo {title} {Lattice quantum electrodynamics in (3+1)-dimensions at
  finite density with tensor networks},\ }\href
  {https://doi.org/10.1038/s41467-021-23646-3} {\bibfield  {journal} {\bibinfo
  {journal} {Nature Communications}\ }\textbf {\bibinfo {volume} {12}},\
  \bibinfo {pages} {3600} (\bibinfo {year} {2021})}\BibitemShut {NoStop}%
\bibitem [{\citenamefont {Emonts}\ \emph {et~al.}(2023)\citenamefont {Emonts},
  \citenamefont {Kelman}, \citenamefont {Borla}, \citenamefont {Moroz},
  \citenamefont {Gazit},\ and\ \citenamefont {Zohar}}]{emonts2022z2}%
  \BibitemOpen
  \bibfield  {author} {\bibinfo {author} {\bibfnamefont {P.}~\bibnamefont
  {Emonts}}, \bibinfo {author} {\bibfnamefont {A.}~\bibnamefont {Kelman}},
  \bibinfo {author} {\bibfnamefont {U.}~\bibnamefont {Borla}}, \bibinfo
  {author} {\bibfnamefont {S.}~\bibnamefont {Moroz}}, \bibinfo {author}
  {\bibfnamefont {S.}~\bibnamefont {Gazit}},\ and\ \bibinfo {author}
  {\bibfnamefont {E.}~\bibnamefont {Zohar}},\ }\bibfield  {title} {\bibinfo
  {title} {{Finding the ground state of a lattice gauge theory with fermionic
  tensor networks: A $2+1\mathrm{D}$ ${\mathbb{Z}}_{2}$ demonstration}},\
  }\href {https://doi.org/10.1103/PhysRevD.107.014505} {\bibfield  {journal}
  {\bibinfo  {journal} {Phys. Rev. D}\ }\textbf {\bibinfo {volume} {107}},\
  \bibinfo {pages} {014505} (\bibinfo {year} {2023})}\BibitemShut {NoStop}%
\bibitem [{\citenamefont {Tong}\ \emph {et~al.}(2022)\citenamefont {Tong},
  \citenamefont {Albert}, \citenamefont {McClean}, \citenamefont {Preskill},\
  and\ \citenamefont {Su}}]{tong2022provablyaccurate}%
  \BibitemOpen
  \bibfield  {author} {\bibinfo {author} {\bibfnamefont {Y.}~\bibnamefont
  {Tong}}, \bibinfo {author} {\bibfnamefont {V.~V.}\ \bibnamefont {Albert}},
  \bibinfo {author} {\bibfnamefont {J.~R.}\ \bibnamefont {McClean}}, \bibinfo
  {author} {\bibfnamefont {J.}~\bibnamefont {Preskill}},\ and\ \bibinfo
  {author} {\bibfnamefont {Y.}~\bibnamefont {Su}},\ }\bibfield  {title}
  {\bibinfo {title} {Provably accurate simulation of gauge theories and bosonic
  systems},\ }\href {https://doi.org/10.22331/q-2022-09-22-816} {\bibfield
  {journal} {\bibinfo  {journal} {{Quantum}}\ }\textbf {\bibinfo {volume}
  {6}},\ \bibinfo {pages} {816} (\bibinfo {year} {2022})}\BibitemShut {NoStop}%
\bibitem [{\citenamefont {Buyens}\ \emph {et~al.}(2014)\citenamefont {Buyens},
  \citenamefont {Haegeman}, \citenamefont {Van~Acoleyen}, \citenamefont
  {Verschelde},\ and\ \citenamefont {Verstraete}}]{buyens2014}%
  \BibitemOpen
  \bibfield  {author} {\bibinfo {author} {\bibfnamefont {B.}~\bibnamefont
  {Buyens}}, \bibinfo {author} {\bibfnamefont {J.}~\bibnamefont {Haegeman}},
  \bibinfo {author} {\bibfnamefont {K.}~\bibnamefont {Van~Acoleyen}}, \bibinfo
  {author} {\bibfnamefont {H.}~\bibnamefont {Verschelde}},\ and\ \bibinfo
  {author} {\bibfnamefont {F.}~\bibnamefont {Verstraete}},\ }\bibfield  {title}
  {\bibinfo {title} {{Matrix Product States for Gauge Field Theories}},\ }\href
  {https://doi.org/10.1103/PhysRevLett.113.091601} {\bibfield  {journal}
  {\bibinfo  {journal} {Phys. Rev. Lett.}\ }\textbf {\bibinfo {volume} {113}},\
  \bibinfo {pages} {091601} (\bibinfo {year} {2014})}\BibitemShut {NoStop}%
\bibitem [{\citenamefont {K{\"u}hn}\ \emph {et~al.}(2015)\citenamefont
  {K{\"u}hn}, \citenamefont {Zohar}, \citenamefont {Cirac},\ and\ \citenamefont
  {Ba{\~n}uls}}]{kuehn2015string}%
  \BibitemOpen
  \bibfield  {author} {\bibinfo {author} {\bibfnamefont {S.}~\bibnamefont
  {K{\"u}hn}}, \bibinfo {author} {\bibfnamefont {E.}~\bibnamefont {Zohar}},
  \bibinfo {author} {\bibfnamefont {J.~I.}\ \bibnamefont {Cirac}},\ and\
  \bibinfo {author} {\bibfnamefont {M.~C.}\ \bibnamefont {Ba{\~n}uls}},\
  }\bibfield  {title} {\bibinfo {title} {{Non-Abelian string breaking phenomena
  with matrix product states}},\ }\href
  {https://doi.org/https://doi.org/10.1007/JHEP07%282015%29130} {\bibfield
  {journal} {\bibinfo  {journal} {Journal of High Energy Physics}\ }\textbf
  {\bibinfo {volume} {2015}},\ \bibinfo {pages} {130} (\bibinfo {year}
  {2015})}\BibitemShut {NoStop}%
\bibitem [{\citenamefont {Buyens}\ \emph {et~al.}(2016)\citenamefont {Buyens},
  \citenamefont {Haegeman}, \citenamefont {Verschelde}, \citenamefont
  {Verstraete},\ and\ \citenamefont {Van~Acoleyen}}]{buyens2016}%
  \BibitemOpen
  \bibfield  {author} {\bibinfo {author} {\bibfnamefont {B.}~\bibnamefont
  {Buyens}}, \bibinfo {author} {\bibfnamefont {J.}~\bibnamefont {Haegeman}},
  \bibinfo {author} {\bibfnamefont {H.}~\bibnamefont {Verschelde}}, \bibinfo
  {author} {\bibfnamefont {F.}~\bibnamefont {Verstraete}},\ and\ \bibinfo
  {author} {\bibfnamefont {K.}~\bibnamefont {Van~Acoleyen}},\ }\bibfield
  {title} {\bibinfo {title} {{Confinement and String Breaking for
  ${\mathrm{QED}}_{2}$ in the Hamiltonian Picture}},\ }\href
  {https://doi.org/10.1103/PhysRevX.6.041040} {\bibfield  {journal} {\bibinfo
  {journal} {Phys. Rev. X}\ }\textbf {\bibinfo {volume} {6}},\ \bibinfo {pages}
  {041040} (\bibinfo {year} {2016})}\BibitemShut {NoStop}%
\bibitem [{\citenamefont {Pichler}\ \emph {et~al.}(2016)\citenamefont
  {Pichler}, \citenamefont {Dalmonte}, \citenamefont {Rico}, \citenamefont
  {Zoller},\ and\ \citenamefont {Montangero}}]{pichler2016realt}%
  \BibitemOpen
  \bibfield  {author} {\bibinfo {author} {\bibfnamefont {T.}~\bibnamefont
  {Pichler}}, \bibinfo {author} {\bibfnamefont {M.}~\bibnamefont {Dalmonte}},
  \bibinfo {author} {\bibfnamefont {E.}~\bibnamefont {Rico}}, \bibinfo {author}
  {\bibfnamefont {P.}~\bibnamefont {Zoller}},\ and\ \bibinfo {author}
  {\bibfnamefont {S.}~\bibnamefont {Montangero}},\ }\bibfield  {title}
  {\bibinfo {title} {{Real-Time Dynamics in U(1) Lattice Gauge Theories with
  Tensor Networks}},\ }\href {https://doi.org/10.1103/PhysRevX.6.011023}
  {\bibfield  {journal} {\bibinfo  {journal} {Phys. Rev. X}\ }\textbf {\bibinfo
  {volume} {6}},\ \bibinfo {pages} {011023} (\bibinfo {year}
  {2016})}\BibitemShut {NoStop}%
\bibitem [{\citenamefont {Chanda}\ \emph {et~al.}(2020)\citenamefont {Chanda},
  \citenamefont {Zakrzewski}, \citenamefont {Lewenstein},\ and\ \citenamefont
  {Tagliacozzo}}]{chanda2020quenches}%
  \BibitemOpen
  \bibfield  {author} {\bibinfo {author} {\bibfnamefont {T.}~\bibnamefont
  {Chanda}}, \bibinfo {author} {\bibfnamefont {J.}~\bibnamefont {Zakrzewski}},
  \bibinfo {author} {\bibfnamefont {M.}~\bibnamefont {Lewenstein}},\ and\
  \bibinfo {author} {\bibfnamefont {L.}~\bibnamefont {Tagliacozzo}},\
  }\bibfield  {title} {\bibinfo {title} {{Confinement and Lack of
  Thermalization after Quenches in the Bosonic Schwinger Model}},\ }\href
  {https://doi.org/10.1103/PhysRevLett.124.180602} {\bibfield  {journal}
  {\bibinfo  {journal} {Phys. Rev. Lett.}\ }\textbf {\bibinfo {volume} {124}},\
  \bibinfo {pages} {180602} (\bibinfo {year} {2020})}\BibitemShut {NoStop}%
\bibitem [{\citenamefont {Notarnicola}\ \emph {et~al.}(2020)\citenamefont
  {Notarnicola}, \citenamefont {Collura},\ and\ \citenamefont
  {Montangero}}]{notarnicola2020ryd}%
  \BibitemOpen
  \bibfield  {author} {\bibinfo {author} {\bibfnamefont {S.}~\bibnamefont
  {Notarnicola}}, \bibinfo {author} {\bibfnamefont {M.}~\bibnamefont
  {Collura}},\ and\ \bibinfo {author} {\bibfnamefont {S.}~\bibnamefont
  {Montangero}},\ }\bibfield  {title} {\bibinfo {title} {{Real-time-dynamics
  quantum simulation of $(1+1)\text{-dimensional}$ lattice QED with Rydberg
  atoms}},\ }\href {https://doi.org/10.1103/PhysRevResearch.2.013288}
  {\bibfield  {journal} {\bibinfo  {journal} {Phys. Rev. Res.}\ }\textbf
  {\bibinfo {volume} {2}},\ \bibinfo {pages} {013288} (\bibinfo {year}
  {2020})}\BibitemShut {NoStop}%
\bibitem [{\citenamefont {Rigobello}\ \emph {et~al.}(2021)\citenamefont
  {Rigobello}, \citenamefont {Notarnicola}, \citenamefont {Magnifico},\ and\
  \citenamefont {Montangero}}]{rigobello2021entanglement}%
  \BibitemOpen
  \bibfield  {author} {\bibinfo {author} {\bibfnamefont {M.}~\bibnamefont
  {Rigobello}}, \bibinfo {author} {\bibfnamefont {S.}~\bibnamefont
  {Notarnicola}}, \bibinfo {author} {\bibfnamefont {G.}~\bibnamefont
  {Magnifico}},\ and\ \bibinfo {author} {\bibfnamefont {S.}~\bibnamefont
  {Montangero}},\ }\bibfield  {title} {\bibinfo {title} {{Entanglement
  generation in $(1+1)\mathrm{D}$ QED scattering processes}},\ }\href
  {https://doi.org/10.1103/PhysRevD.104.114501} {\bibfield  {journal} {\bibinfo
   {journal} {Phys. Rev. D}\ }\textbf {\bibinfo {volume} {104}},\ \bibinfo
  {pages} {114501} (\bibinfo {year} {2021})}\BibitemShut {NoStop}%
\bibitem [{\citenamefont {Belyansky}\ \emph {et~al.}(2023)\citenamefont
  {Belyansky}, \citenamefont {Whitsitt}, \citenamefont {Mueller}, \citenamefont
  {Fahimniya}, \citenamefont {Bennewitz}, \citenamefont {Davoudi},\ and\
  \citenamefont {Gorshkov}}]{belyansky2023high}%
  \BibitemOpen
  \bibfield  {author} {\bibinfo {author} {\bibfnamefont {R.}~\bibnamefont
  {Belyansky}}, \bibinfo {author} {\bibfnamefont {S.}~\bibnamefont {Whitsitt}},
  \bibinfo {author} {\bibfnamefont {N.}~\bibnamefont {Mueller}}, \bibinfo
  {author} {\bibfnamefont {A.}~\bibnamefont {Fahimniya}}, \bibinfo {author}
  {\bibfnamefont {E.~R.}\ \bibnamefont {Bennewitz}}, \bibinfo {author}
  {\bibfnamefont {Z.}~\bibnamefont {Davoudi}},\ and\ \bibinfo {author}
  {\bibfnamefont {A.~V.}\ \bibnamefont {Gorshkov}},\ }\bibfield  {title}
  {\bibinfo {title} {{High-Energy Collision of Quarks and Hadrons in the
  Schwinger Model: From Tensor Networks to Circuit QED}},\ }\bibfield
  {journal} {\bibinfo  {journal} {arXiv preprint}\ }\href
  {https://doi.org/10.48550/arXiv.2307.02522} {10.48550/arXiv.2307.02522}
  (\bibinfo {year} {2023})\BibitemShut {NoStop}%
\bibitem [{\citenamefont {Kormos}\ \emph {et~al.}(2017)\citenamefont {Kormos},
  \citenamefont {Collura}, \citenamefont {Tak{\'a}cs},\ and\ \citenamefont
  {Calabrese}}]{kormos2017real}%
  \BibitemOpen
  \bibfield  {author} {\bibinfo {author} {\bibfnamefont {M.}~\bibnamefont
  {Kormos}}, \bibinfo {author} {\bibfnamefont {M.}~\bibnamefont {Collura}},
  \bibinfo {author} {\bibfnamefont {G.}~\bibnamefont {Tak{\'a}cs}},\ and\
  \bibinfo {author} {\bibfnamefont {P.}~\bibnamefont {Calabrese}},\ }\bibfield
  {title} {\bibinfo {title} {Real-time confinement following a quantum quench
  to a non-integrable model},\ }\href
  {https://doi.org/https://doi.org/10.1038/nphys3934} {\bibfield  {journal}
  {\bibinfo  {journal} {Nature Physics}\ }\textbf {\bibinfo {volume} {13}},\
  \bibinfo {pages} {246} (\bibinfo {year} {2017})}\BibitemShut {NoStop}%
\bibitem [{\citenamefont {Vovrosh}\ \emph
  {et~al.}(2022{\natexlab{a}})\citenamefont {Vovrosh}, \citenamefont {Zhao},
  \citenamefont {Knolle},\ and\ \citenamefont
  {Bastianello}}]{vovrosh2022confinement}%
  \BibitemOpen
  \bibfield  {author} {\bibinfo {author} {\bibfnamefont {J.}~\bibnamefont
  {Vovrosh}}, \bibinfo {author} {\bibfnamefont {H.}~\bibnamefont {Zhao}},
  \bibinfo {author} {\bibfnamefont {J.}~\bibnamefont {Knolle}},\ and\ \bibinfo
  {author} {\bibfnamefont {A.}~\bibnamefont {Bastianello}},\ }\bibfield
  {title} {\bibinfo {title} {Confinement-induced impurity states in spin
  chains},\ }\href
  {https://doi.org/https://doi.org/10.1103/PhysRevB.105.L100301} {\bibfield
  {journal} {\bibinfo  {journal} {Phys. Rev. B}\ }\textbf {\bibinfo {volume}
  {105}},\ \bibinfo {pages} {L100301} (\bibinfo {year}
  {2022}{\natexlab{a}})}\BibitemShut {NoStop}%
\bibitem [{\citenamefont {Van~Damme}\ \emph {et~al.}(2021)\citenamefont
  {Van~Damme}, \citenamefont {Vanderstraeten}, \citenamefont {De~Nardis},
  \citenamefont {Haegeman},\ and\ \citenamefont
  {Verstraete}}]{vandamme2021scat}%
  \BibitemOpen
  \bibfield  {author} {\bibinfo {author} {\bibfnamefont {M.}~\bibnamefont
  {Van~Damme}}, \bibinfo {author} {\bibfnamefont {L.}~\bibnamefont
  {Vanderstraeten}}, \bibinfo {author} {\bibfnamefont {J.}~\bibnamefont
  {De~Nardis}}, \bibinfo {author} {\bibfnamefont {J.}~\bibnamefont
  {Haegeman}},\ and\ \bibinfo {author} {\bibfnamefont {F.}~\bibnamefont
  {Verstraete}},\ }\bibfield  {title} {\bibinfo {title} {Real-time scattering
  of interacting quasiparticles in quantum spin chains},\ }\href
  {https://doi.org/10.1103/PhysRevResearch.3.013078} {\bibfield  {journal}
  {\bibinfo  {journal} {Phys. Rev. Res.}\ }\textbf {\bibinfo {volume} {3}},\
  \bibinfo {pages} {013078} (\bibinfo {year} {2021})}\BibitemShut {NoStop}%
\bibitem [{\citenamefont {Surace}\ and\ \citenamefont
  {Lerose}(2021)}]{surace2021scattering}%
  \BibitemOpen
  \bibfield  {author} {\bibinfo {author} {\bibfnamefont {F.~M.}\ \bibnamefont
  {Surace}}\ and\ \bibinfo {author} {\bibfnamefont {A.}~\bibnamefont
  {Lerose}},\ }\bibfield  {title} {\bibinfo {title} {Scattering of mesons in
  quantum simulators},\ }\href
  {https://doi.org/https://doi.org/10.1088/1367-2630/abfc40} {\bibfield
  {journal} {\bibinfo  {journal} {New Journal of Physics}\ }\textbf {\bibinfo
  {volume} {23}},\ \bibinfo {pages} {062001} (\bibinfo {year}
  {2021})}\BibitemShut {NoStop}%
\bibitem [{\citenamefont {Karpov}\ \emph {et~al.}(2022)\citenamefont {Karpov},
  \citenamefont {Zhu}, \citenamefont {Heller},\ and\ \citenamefont
  {Heyl}}]{karpov2022spatiotemporal}%
  \BibitemOpen
  \bibfield  {author} {\bibinfo {author} {\bibfnamefont {P.}~\bibnamefont
  {Karpov}}, \bibinfo {author} {\bibfnamefont {G.-Y.}\ \bibnamefont {Zhu}},
  \bibinfo {author} {\bibfnamefont {M.~P.}\ \bibnamefont {Heller}},\ and\
  \bibinfo {author} {\bibfnamefont {M.}~\bibnamefont {Heyl}},\ }\bibfield
  {title} {\bibinfo {title} {Spatiotemporal dynamics of particle collisions in
  quantum spin chains},\ }\href
  {https://doi.org/10.1103/PhysRevResearch.4.L032001} {\bibfield  {journal}
  {\bibinfo  {journal} {Phys. Rev. Res.}\ }\textbf {\bibinfo {volume} {4}},\
  \bibinfo {pages} {L032001} (\bibinfo {year} {2022})}\BibitemShut {NoStop}%
\bibitem [{\citenamefont {Milsted}\ \emph {et~al.}(2022)\citenamefont
  {Milsted}, \citenamefont {Liu}, \citenamefont {Preskill},\ and\ \citenamefont
  {Vidal}}]{milsted2022collisions}%
  \BibitemOpen
  \bibfield  {author} {\bibinfo {author} {\bibfnamefont {A.}~\bibnamefont
  {Milsted}}, \bibinfo {author} {\bibfnamefont {J.}~\bibnamefont {Liu}},
  \bibinfo {author} {\bibfnamefont {J.}~\bibnamefont {Preskill}},\ and\
  \bibinfo {author} {\bibfnamefont {G.}~\bibnamefont {Vidal}},\ }\bibfield
  {title} {\bibinfo {title} {Collisions of false-vacuum bubble walls in a
  quantum spin chain},\ }\href
  {https://doi.org/https://doi.org/10.1103/PRXQuantum.3.020316} {\bibfield
  {journal} {\bibinfo  {journal} {PRX Quantum}\ }\textbf {\bibinfo {volume}
  {3}},\ \bibinfo {pages} {020316} (\bibinfo {year} {2022})}\BibitemShut
  {NoStop}%
\bibitem [{\citenamefont {Vovrosh}\ \emph
  {et~al.}(2022{\natexlab{b}})\citenamefont {Vovrosh}, \citenamefont
  {Mukherjee}, \citenamefont {Bastianello},\ and\ \citenamefont
  {Knolle}}]{vovrosh2022hadronformation}%
  \BibitemOpen
  \bibfield  {author} {\bibinfo {author} {\bibfnamefont {J.}~\bibnamefont
  {Vovrosh}}, \bibinfo {author} {\bibfnamefont {R.}~\bibnamefont {Mukherjee}},
  \bibinfo {author} {\bibfnamefont {A.}~\bibnamefont {Bastianello}},\ and\
  \bibinfo {author} {\bibfnamefont {J.}~\bibnamefont {Knolle}},\ }\bibfield
  {title} {\bibinfo {title} {{Dynamical Hadron Formation in Long-Range
  Interacting Quantum Spin Chains}},\ }\href
  {https://doi.org/10.1103/PRXQuantum.3.040309} {\bibfield  {journal} {\bibinfo
   {journal} {PRX Quantum}\ }\textbf {\bibinfo {volume} {3}},\ \bibinfo {pages}
  {040309} (\bibinfo {year} {2022}{\natexlab{b}})}\BibitemShut {NoStop}%
\bibitem [{\citenamefont {Chai}\ \emph {et~al.}(2023)\citenamefont {Chai},
  \citenamefont {Crippa}, \citenamefont {Jansen}, \citenamefont {K\"uhn},
  \citenamefont {Pascuzzi}, \citenamefont {Tacchino},\ and\ \citenamefont
  {Tavernelli}}]{chai2023thirring}%
  \BibitemOpen
  \bibfield  {author} {\bibinfo {author} {\bibfnamefont {Y.}~\bibnamefont
  {Chai}}, \bibinfo {author} {\bibfnamefont {A.}~\bibnamefont {Crippa}},
  \bibinfo {author} {\bibfnamefont {K.}~\bibnamefont {Jansen}}, \bibinfo
  {author} {\bibfnamefont {S.}~\bibnamefont {K\"uhn}}, \bibinfo {author}
  {\bibfnamefont {V.~R.}\ \bibnamefont {Pascuzzi}}, \bibinfo {author}
  {\bibfnamefont {F.}~\bibnamefont {Tacchino}},\ and\ \bibinfo {author}
  {\bibfnamefont {I.}~\bibnamefont {Tavernelli}},\ }\bibfield  {title}
  {\bibinfo {title} {Entanglement production from scattering of fermionic wave
  packets: a quantum computing approach},\ }\href@noop {} {\bibfield  {journal}
  {\bibinfo  {journal} {arXiv:2312.02272}\ } (\bibinfo {year} {2023})},\
  \Eprint {https://arxiv.org/abs/2312.02272} {arXiv:2312.02272 [quant-ph]}
  \BibitemShut {NoStop}%
\bibitem [{\citenamefont {Farrell}\ \emph {et~al.}(2024)\citenamefont
  {Farrell}, \citenamefont {Illa}, \citenamefont {Ciavarella},\ and\
  \citenamefont {Savage}}]{farrell2024quantum}%
  \BibitemOpen
  \bibfield  {author} {\bibinfo {author} {\bibfnamefont {R.~C.}\ \bibnamefont
  {Farrell}}, \bibinfo {author} {\bibfnamefont {M.}~\bibnamefont {Illa}},
  \bibinfo {author} {\bibfnamefont {A.~N.}\ \bibnamefont {Ciavarella}},\ and\
  \bibinfo {author} {\bibfnamefont {M.~J.}\ \bibnamefont {Savage}},\
  }\href@noop {} {\bibinfo {title} {{Quantum Simulations of Hadron Dynamics in
  the Schwinger Model using 112 Qubits}}} (\bibinfo {year} {2024}),\ \Eprint
  {https://arxiv.org/abs/2401.08044} {arXiv:2401.08044} \BibitemShut {NoStop}%
\bibitem [{\citenamefont {Shaw}\ \emph {et~al.}(2020)\citenamefont {Shaw},
  \citenamefont {Lougovski}, \citenamefont {Stryker},\ and\ \citenamefont
  {Wiebe}}]{Shaw2020quantumalgorithms}%
  \BibitemOpen
  \bibfield  {author} {\bibinfo {author} {\bibfnamefont {A.~F.}\ \bibnamefont
  {Shaw}}, \bibinfo {author} {\bibfnamefont {P.}~\bibnamefont {Lougovski}},
  \bibinfo {author} {\bibfnamefont {J.~R.}\ \bibnamefont {Stryker}},\ and\
  \bibinfo {author} {\bibfnamefont {N.}~\bibnamefont {Wiebe}},\ }\bibfield
  {title} {\bibinfo {title} {Quantum {A}lgorithms for {S}imulating the
  {L}attice {S}chwinger {M}odel},\ }\href
  {https://doi.org/10.22331/q-2020-08-10-306} {\bibfield  {journal} {\bibinfo
  {journal} {{Quantum}}\ }\textbf {\bibinfo {volume} {4}},\ \bibinfo {pages}
  {306} (\bibinfo {year} {2020})}\BibitemShut {NoStop}%
\bibitem [{\citenamefont {Schwinger}(1962{\natexlab{a}})}]{schwinger1962gauge}%
  \BibitemOpen
  \bibfield  {author} {\bibinfo {author} {\bibfnamefont {J.}~\bibnamefont
  {Schwinger}},\ }\bibfield  {title} {\bibinfo {title} {{Gauge Invariance and
  Mass. II}},\ }\href {https://doi.org/10.1103/PhysRev.128.2425} {\bibfield
  {journal} {\bibinfo  {journal} {Phys. Rev.}\ }\textbf {\bibinfo {volume}
  {128}},\ \bibinfo {pages} {2425} (\bibinfo {year}
  {1962}{\natexlab{a}})}\BibitemShut {NoStop}%
\bibitem [{\citenamefont {Coleman}(1976)}]{coleman1976more}%
  \BibitemOpen
  \bibfield  {author} {\bibinfo {author} {\bibfnamefont {S.~R.}\ \bibnamefont
  {Coleman}},\ }\bibfield  {title} {\bibinfo {title} {{More About the Massive
  Schwinger Model}},\ }\href {https://doi.org/10.1016/0003-4916(76)90280-3}
  {\bibfield  {journal} {\bibinfo  {journal} {Annals Phys.}\ }\textbf {\bibinfo
  {volume} {101}},\ \bibinfo {pages} {239} (\bibinfo {year}
  {1976})}\BibitemShut {NoStop}%
\bibitem [{ham(1982)}]{hamer1982}%
  \BibitemOpen
  \bibfield  {title} {\bibinfo {title} {{The massive Schwinger model on a
  lattice: Background field, chiral symmetry and the string tension}},\ }\href
  {https://doi.org/https://doi.org/10.1016/0550-3213(82)90229-2} {\bibfield
  {journal} {\bibinfo  {journal} {Nuclear Physics B}\ }\textbf {\bibinfo
  {volume} {208}},\ \bibinfo {pages} {413} (\bibinfo {year}
  {1982})}\BibitemShut {NoStop}%
\bibitem [{\citenamefont {Hamer}\ \emph {et~al.}(1997)\citenamefont {Hamer},
  \citenamefont {Zheng},\ and\ \citenamefont {Oitmaa}}]{hamer1997series}%
  \BibitemOpen
  \bibfield  {author} {\bibinfo {author} {\bibfnamefont {C.~J.}\ \bibnamefont
  {Hamer}}, \bibinfo {author} {\bibfnamefont {W.-h.}\ \bibnamefont {Zheng}},\
  and\ \bibinfo {author} {\bibfnamefont {J.}~\bibnamefont {Oitmaa}},\
  }\bibfield  {title} {\bibinfo {title} {{{Series expansions for the massive
  Schwinger model in Hamiltonian lattice theory}}},\ }\href
  {https://doi.org/10.1103/PhysRevD.56.55} {\bibfield  {journal} {\bibinfo
  {journal} {Phys. Rev. D}\ }\textbf {\bibinfo {volume} {56}},\ \bibinfo
  {pages} {55} (\bibinfo {year} {1997})},\ \Eprint
  {https://arxiv.org/abs/hep-lat/9701015} {arXiv:hep-lat/9701015} \BibitemShut
  {NoStop}%
\bibitem [{\citenamefont {Byrnes}\ \emph {et~al.}(2002)\citenamefont {Byrnes},
  \citenamefont {Sriganesh}, \citenamefont {Bursill},\ and\ \citenamefont
  {Hamer}}]{Byrnes:2002nv}%
  \BibitemOpen
  \bibfield  {author} {\bibinfo {author} {\bibfnamefont {T.}~\bibnamefont
  {Byrnes}}, \bibinfo {author} {\bibfnamefont {P.}~\bibnamefont {Sriganesh}},
  \bibinfo {author} {\bibfnamefont {R.~J.}\ \bibnamefont {Bursill}},\ and\
  \bibinfo {author} {\bibfnamefont {C.~J.}\ \bibnamefont {Hamer}},\ }\bibfield
  {title} {\bibinfo {title} {{{Density matrix renormalization group approach to
  the massive Schwinger model}}},\ }\href
  {https://doi.org/10.1103/PhysRevD.66.013002} {\bibfield  {journal} {\bibinfo
  {journal} {Phys. Rev. D}\ }\textbf {\bibinfo {volume} {66}},\ \bibinfo
  {pages} {013002} (\bibinfo {year} {2002})},\ \Eprint
  {https://arxiv.org/abs/hep-lat/0202014} {arXiv:hep-lat/0202014} \BibitemShut
  {NoStop}%
\bibitem [{\citenamefont {Schwinger}(1962{\natexlab{b}})}]{schwinger1}%
  \BibitemOpen
  \bibfield  {author} {\bibinfo {author} {\bibfnamefont {J.}~\bibnamefont
  {Schwinger}},\ }\bibfield  {title} {\bibinfo {title} {{Gauge Invariance and
  Mass}},\ }\href {https://doi.org/10.1103/PhysRev.125.397} {\bibfield
  {journal} {\bibinfo  {journal} {Phys. Rev.}\ }\textbf {\bibinfo {volume}
  {125}},\ \bibinfo {pages} {397} (\bibinfo {year}
  {1962}{\natexlab{b}})}\BibitemShut {NoStop}%
\bibitem [{\citenamefont {Adam}(1997)}]{adam1997}%
  \BibitemOpen
  \bibfield  {author} {\bibinfo {author} {\bibfnamefont {C.}~\bibnamefont
  {Adam}},\ }\bibfield  {title} {\bibinfo {title} {{Scattering processes in the
  massive Schwinger model}},\ }\href {https://doi.org/10.1103/PhysRevD.55.6299}
  {\bibfield  {journal} {\bibinfo  {journal} {Phys. Rev. D}\ }\textbf {\bibinfo
  {volume} {55}},\ \bibinfo {pages} {6299} (\bibinfo {year}
  {1997})}\BibitemShut {NoStop}%
\bibitem [{\citenamefont {Coleman}\ \emph {et~al.}(1975)\citenamefont
  {Coleman}, \citenamefont {Jackiw},\ and\ \citenamefont
  {Susskind}}]{coleman75}%
  \BibitemOpen
  \bibfield  {author} {\bibinfo {author} {\bibfnamefont {S.}~\bibnamefont
  {Coleman}}, \bibinfo {author} {\bibfnamefont {R.}~\bibnamefont {Jackiw}},\
  and\ \bibinfo {author} {\bibfnamefont {L.}~\bibnamefont {Susskind}},\
  }\bibfield  {title} {\bibinfo {title} {Charge shielding and quark confinement
  in the massive schwinger model},\ }\href
  {https://doi.org/https://doi.org/10.1016/0003-4916(75)90212-2} {\bibfield
  {journal} {\bibinfo  {journal} {Annals of Physics}\ }\textbf {\bibinfo
  {volume} {93}},\ \bibinfo {pages} {267} (\bibinfo {year} {1975})}\BibitemShut
  {NoStop}%
\bibitem [{\citenamefont {Adam}(1996)}]{ADAM1996383}%
  \BibitemOpen
  \bibfield  {author} {\bibinfo {author} {\bibfnamefont {C.}~\bibnamefont
  {Adam}},\ }\bibfield  {title} {\bibinfo {title} {{The Schwinger mass in the
  massive Schwinger model}},\ }\href
  {https://doi.org/https://doi.org/10.1016/0370-2693(96)00695-8} {\bibfield
  {journal} {\bibinfo  {journal} {Physics Letters B}\ }\textbf {\bibinfo
  {volume} {382}},\ \bibinfo {pages} {383} (\bibinfo {year}
  {1996})}\BibitemShut {NoStop}%
\bibitem [{\citenamefont {Adam}(2003)}]{ADAM2003}%
  \BibitemOpen
  \bibfield  {author} {\bibinfo {author} {\bibfnamefont {C.}~\bibnamefont
  {Adam}},\ }\bibfield  {title} {\bibinfo {title} {{Improved vector and scalar
  masses in the massive Schwinger model}},\ }\href
  {https://doi.org/https://doi.org/10.1016/S0370-2693(03)00051-0} {\bibfield
  {journal} {\bibinfo  {journal} {Physics Letters B}\ }\textbf {\bibinfo
  {volume} {555}},\ \bibinfo {pages} {132} (\bibinfo {year}
  {2003})}\BibitemShut {NoStop}%
\bibitem [{\citenamefont {Mo}\ and\ \citenamefont {Perry}(1993)}]{mo1993basis}%
  \BibitemOpen
  \bibfield  {author} {\bibinfo {author} {\bibfnamefont {Y.}~\bibnamefont
  {Mo}}\ and\ \bibinfo {author} {\bibfnamefont {R.~J.}\ \bibnamefont {Perry}},\
  }\bibfield  {title} {\bibinfo {title} {{Basis Function Calculations for the
  Massive Schwinger Model in the Light-Front Tamm-Dancoff Approximation}},\
  }\href {https://doi.org/https://doi.org/10.1006/jcph.1993.1171} {\bibfield
  {journal} {\bibinfo  {journal} {Journal of Computational Physics}\ }\textbf
  {\bibinfo {volume} {108}},\ \bibinfo {pages} {159} (\bibinfo {year}
  {1993})}\BibitemShut {NoStop}%
\bibitem [{\citenamefont {Harada}\ \emph {et~al.}(1995)\citenamefont {Harada},
  \citenamefont {Okazaki},\ and\ \citenamefont {Taniguchi}}]{harada1995}%
  \BibitemOpen
  \bibfield  {author} {\bibinfo {author} {\bibfnamefont {K.}~\bibnamefont
  {Harada}}, \bibinfo {author} {\bibfnamefont {A.}~\bibnamefont {Okazaki}},\
  and\ \bibinfo {author} {\bibfnamefont {M.-a.}\ \bibnamefont {Taniguchi}},\
  }\bibfield  {title} {\bibinfo {title} {{Six-body light-front Tamm-Dancoff
  approximation and wave functions for the massive Schwinger model}},\ }\href
  {https://doi.org/10.1103/PhysRevD.52.2429} {\bibfield  {journal} {\bibinfo
  {journal} {Phys. Rev. D}\ }\textbf {\bibinfo {volume} {52}},\ \bibinfo
  {pages} {2429} (\bibinfo {year} {1995})}\BibitemShut {NoStop}%
\bibitem [{\citenamefont {Kogut}\ and\ \citenamefont
  {Susskind}(1975)}]{kogut1975j}%
  \BibitemOpen
  \bibfield  {author} {\bibinfo {author} {\bibfnamefont {J.}~\bibnamefont
  {Kogut}}\ and\ \bibinfo {author} {\bibfnamefont {L.}~\bibnamefont
  {Susskind}},\ }\bibfield  {title} {\bibinfo {title} {{Hamiltonian formulation
  of Wilson's lattice gauge theories}},\ }\href
  {https://doi.org/10.1103/PhysRevD.11.395} {\bibfield  {journal} {\bibinfo
  {journal} {Phys. Rev. D}\ }\textbf {\bibinfo {volume} {11}},\ \bibinfo
  {pages} {395} (\bibinfo {year} {1975})}\BibitemShut {NoStop}%
\bibitem [{\citenamefont {Banks}\ \emph {et~al.}(1976)\citenamefont {Banks},
  \citenamefont {Susskind},\ and\ \citenamefont {Kogut}}]{banks1976strong}%
  \BibitemOpen
  \bibfield  {author} {\bibinfo {author} {\bibfnamefont {T.}~\bibnamefont
  {Banks}}, \bibinfo {author} {\bibfnamefont {L.}~\bibnamefont {Susskind}},\
  and\ \bibinfo {author} {\bibfnamefont {J.}~\bibnamefont {Kogut}},\ }\bibfield
   {title} {\bibinfo {title} {{Strong-coupling calculations of lattice gauge
  theories: (1 + 1)-dimensional exercises}},\ }\href
  {https://doi.org/10.1103/PhysRevD.13.1043} {\bibfield  {journal} {\bibinfo
  {journal} {Phys. Rev. D}\ }\textbf {\bibinfo {volume} {13}},\ \bibinfo
  {pages} {1043} (\bibinfo {year} {1976})}\BibitemShut {NoStop}%
\bibitem [{\citenamefont {Ba{\~n}uls}\ \emph {et~al.}(2013)\citenamefont
  {Ba{\~n}uls}, \citenamefont {Cichy}, \citenamefont {Cirac},\ and\
  \citenamefont {Jansen}}]{banuls2013mass}%
  \BibitemOpen
  \bibfield  {author} {\bibinfo {author} {\bibfnamefont {M.~C.}\ \bibnamefont
  {Ba{\~n}uls}}, \bibinfo {author} {\bibfnamefont {K.}~\bibnamefont {Cichy}},
  \bibinfo {author} {\bibfnamefont {J.~I.}\ \bibnamefont {Cirac}},\ and\
  \bibinfo {author} {\bibfnamefont {K.}~\bibnamefont {Jansen}},\ }\bibfield
  {title} {\bibinfo {title} {The mass spectrum of the schwinger model with
  matrix product states},\ }\href
  {https://doi.org/https://doi.org/10.1007/JHEP11%282013%29158} {\bibfield
  {journal} {\bibinfo  {journal} {Journal of High Energy Physics}\ }\textbf
  {\bibinfo {volume} {2013}},\ \bibinfo {pages} {1} (\bibinfo {year}
  {2013})}\BibitemShut {NoStop}%
\bibitem [{\citenamefont {Cichy}\ \emph {et~al.}(2013)\citenamefont {Cichy},
  \citenamefont {Kujawa-Cichy},\ and\ \citenamefont
  {Szyniszewski}}]{CICHY20131666}%
  \BibitemOpen
  \bibfield  {author} {\bibinfo {author} {\bibfnamefont {K.}~\bibnamefont
  {Cichy}}, \bibinfo {author} {\bibfnamefont {A.}~\bibnamefont
  {Kujawa-Cichy}},\ and\ \bibinfo {author} {\bibfnamefont {M.}~\bibnamefont
  {Szyniszewski}},\ }\bibfield  {title} {\bibinfo {title} {{Lattice Hamiltonian
  approach to the massless Schwinger model: Precise extraction of the mass
  gap}},\ }\href {https://doi.org/https://doi.org/10.1016/j.cpc.2013.02.010}
  {\bibfield  {journal} {\bibinfo  {journal} {Computer Physics Communications}\
  }\textbf {\bibinfo {volume} {184}},\ \bibinfo {pages} {1666} (\bibinfo {year}
  {2013})}\BibitemShut {NoStop}%
\bibitem [{\citenamefont {Pirvu}\ \emph {et~al.}(2010)\citenamefont {Pirvu},
  \citenamefont {Murg}, \citenamefont {Cirac},\ and\ \citenamefont
  {Verstraete}}]{Pirvu_2010}%
  \BibitemOpen
  \bibfield  {author} {\bibinfo {author} {\bibfnamefont {B.}~\bibnamefont
  {Pirvu}}, \bibinfo {author} {\bibfnamefont {V.}~\bibnamefont {Murg}},
  \bibinfo {author} {\bibfnamefont {J.~I.}\ \bibnamefont {Cirac}},\ and\
  \bibinfo {author} {\bibfnamefont {F.}~\bibnamefont {Verstraete}},\ }\bibfield
   {title} {\bibinfo {title} {Matrix product operator representations},\ }\href
  {https://doi.org/10.1088/1367-2630/12/2/025012} {\bibfield  {journal}
  {\bibinfo  {journal} {New Journal of Physics}\ }\textbf {\bibinfo {volume}
  {12}},\ \bibinfo {pages} {025012} (\bibinfo {year} {2010})}\BibitemShut
  {NoStop}%
\bibitem [{\citenamefont {Trotter}(1959)}]{Trotter1959OnTP}%
  \BibitemOpen
  \bibfield  {author} {\bibinfo {author} {\bibfnamefont {H.~F.}\ \bibnamefont
  {Trotter}},\ }\bibfield  {title} {\bibinfo {title} {On the product of
  semi-groups of operators}\ }(\bibinfo {year} {1959})\ pp.\ \bibinfo {pages}
  {545--551}\BibitemShut {NoStop}%
\bibitem [{\citenamefont {Suzuki}(1990)}]{Suzuki1990FractalDO}%
  \BibitemOpen
  \bibfield  {author} {\bibinfo {author} {\bibfnamefont {M.}~\bibnamefont
  {Suzuki}},\ }\bibfield  {title} {\bibinfo {title} {{Fractal decomposition of
  exponential operators with applications to many-body theories and Monte Carlo
  simulations}},\ }\href {https://api.semanticscholar.org/CorpusID:123095683}
  {\bibfield  {journal} {\bibinfo  {journal} {Physics Letters A}\ }\textbf
  {\bibinfo {volume} {146}},\ \bibinfo {pages} {319} (\bibinfo {year}
  {1990})}\BibitemShut {NoStop}%
\bibitem [{\citenamefont {Ba\~nuls}\ \emph {et~al.}(2015)\citenamefont
  {Ba\~nuls}, \citenamefont {Cichy}, \citenamefont {Cirac}, \citenamefont
  {Jansen},\ and\ \citenamefont {Saito}}]{thermal2015}%
  \BibitemOpen
  \bibfield  {author} {\bibinfo {author} {\bibfnamefont {M.~C.}\ \bibnamefont
  {Ba\~nuls}}, \bibinfo {author} {\bibfnamefont {K.}~\bibnamefont {Cichy}},
  \bibinfo {author} {\bibfnamefont {J.~I.}\ \bibnamefont {Cirac}}, \bibinfo
  {author} {\bibfnamefont {K.}~\bibnamefont {Jansen}},\ and\ \bibinfo {author}
  {\bibfnamefont {H.}~\bibnamefont {Saito}},\ }\bibfield  {title} {\bibinfo
  {title} {{Thermal evolution of the Schwinger model with matrix product
  operators}},\ }\href {https://doi.org/10.1103/PhysRevD.92.034519} {\bibfield
  {journal} {\bibinfo  {journal} {Phys. Rev. D}\ }\textbf {\bibinfo {volume}
  {92}},\ \bibinfo {pages} {034519} (\bibinfo {year} {2015})}\BibitemShut
  {NoStop}%
\bibitem [{\citenamefont {Ba\~nuls}\ \emph {et~al.}(2016)\citenamefont
  {Ba\~nuls}, \citenamefont {Cichy}, \citenamefont {Jansen},\ and\
  \citenamefont {Saito}}]{chiral2016}%
  \BibitemOpen
  \bibfield  {author} {\bibinfo {author} {\bibfnamefont {M.~C.}\ \bibnamefont
  {Ba\~nuls}}, \bibinfo {author} {\bibfnamefont {K.}~\bibnamefont {Cichy}},
  \bibinfo {author} {\bibfnamefont {K.}~\bibnamefont {Jansen}},\ and\ \bibinfo
  {author} {\bibfnamefont {H.}~\bibnamefont {Saito}},\ }\bibfield  {title}
  {\bibinfo {title} {{Chiral condensate in the Schwinger model with matrix
  product operators}},\ }\href {https://doi.org/10.1103/PhysRevD.93.094512}
  {\bibfield  {journal} {\bibinfo  {journal} {Phys. Rev. D}\ }\textbf {\bibinfo
  {volume} {93}},\ \bibinfo {pages} {094512} (\bibinfo {year}
  {2016})}\BibitemShut {NoStop}%
\bibitem [{\citenamefont {Impertro}\ \emph {et~al.}(2023)\citenamefont
  {Impertro}, \citenamefont {Karch}, \citenamefont {Wienand}, \citenamefont
  {Huh}, \citenamefont {Schweizer}, \citenamefont {Bloch},\ and\ \citenamefont
  {Aidelsburger}}]{impertro2023local}%
  \BibitemOpen
  \bibfield  {author} {\bibinfo {author} {\bibfnamefont {A.}~\bibnamefont
  {Impertro}}, \bibinfo {author} {\bibfnamefont {S.}~\bibnamefont {Karch}},
  \bibinfo {author} {\bibfnamefont {J.~F.}\ \bibnamefont {Wienand}}, \bibinfo
  {author} {\bibfnamefont {S.}~\bibnamefont {Huh}}, \bibinfo {author}
  {\bibfnamefont {C.}~\bibnamefont {Schweizer}}, \bibinfo {author}
  {\bibfnamefont {I.}~\bibnamefont {Bloch}},\ and\ \bibinfo {author}
  {\bibfnamefont {M.}~\bibnamefont {Aidelsburger}},\ }\bibfield  {title}
  {\bibinfo {title} {Local readout and control of current and kinetic energy
  operators in optical lattices},\ }\bibfield  {journal} {\bibinfo  {journal}
  {arXiv preprint}\ }\href {https://doi.org/10.48550/arXiv.2312.13268}
  {10.48550/arXiv.2312.13268} (\bibinfo {year} {2023})\BibitemShut {NoStop}%
\bibitem [{\citenamefont {Su}\ \emph {et~al.}(2024)\citenamefont {Su},
  \citenamefont {Osborne},\ and\ \citenamefont {Halimeh}}]{su2024cold}%
  \BibitemOpen
  \bibfield  {author} {\bibinfo {author} {\bibfnamefont {G.-X.}\ \bibnamefont
  {Su}}, \bibinfo {author} {\bibfnamefont {J.}~\bibnamefont {Osborne}},\ and\
  \bibinfo {author} {\bibfnamefont {J.~C.}\ \bibnamefont {Halimeh}},\
  }\bibfield  {title} {\bibinfo {title} {{A Cold-Atom Particle Collider}},\
  }\bibfield  {journal} {\bibinfo  {journal} {arXiv preprint}\ }\href
  {https://doi.org/10.48550/arXiv.2401.05489} {10.48550/arXiv.2401.05489}
  (\bibinfo {year} {2024})\BibitemShut {NoStop}%
\end{thebibliography}%
\end{document}